\documentclass[11pt]{article}
\input epsf.sty
\pdfoutput=1 
\usepackage{amsmath,amsthm,epsfig}
\usepackage{booktabs}
\usepackage{graphicx}
\usepackage{subfigure}
\usepackage{mathrsfs}
\usepackage{color}
\usepackage{bm}
\usepackage{float}
\usepackage{amsmath,amssymb}
\usepackage{natbib}
\usepackage{multirow}
\usepackage{tabularx}
\usepackage{calc}
\usepackage{geometry}
\usepackage{algpseudocode}
\usepackage{algorithm}
\usepackage{booktabs}
\usepackage{makecell} %mutiple lines in one cell 
\usepackage{hyperref}
\RequirePackage{threeparttable} %for table notes

\newcommand{\ind}{\perp\!\!\!\!\perp} 
\emergencystretch=\maxdimen
\begin{document}
\title{{\bf A tutorial on optimal dynamic treatment regimes}}

\date{}
\author{Chunyu Wang \\
{chunyu.wang@mrc-bsu.cam.ac.uk}\\
{MRC Biostatistics Unit, University of Cambridge}\\
Brian DM Tom \\ {brian.tom@mrc-bsu.cam.ac.uk} \\ {MRC Biostatistics Unit, University of Cambridge}}

\maketitle

\begin{center}
\begin{minipage}{140mm}
\begin{center}{\bf Abstract}
\end{center}
A dynamic treatment regime (DTR) is a sequence of treatment decision rules tailored to an individual's evolving status over time. In precision medicine, much focus has been placed on finding an optimal DTR which, if followed by everyone in the population, would yield the best outcome on average; and extensive investigations have been conducted from both methodological and applied standpoints. The purpose of this tutorial is to provide readers who are interested in optimal DTRs with a systematic, detailed, but accessible introduction, including the formal definition and formulation of this topic within the framework of causal inference, identification assumptions required to link the causal quantity of interest to the observed data, existing statistical models and estimation methods for learning the optimal regime from the data, and application of these methods to both simulated and real data. 
\end{minipage}\end{center}

\section{Introduction}\label{sec:intro}
Motivated by precision medicine which attempts to target the right treatment to the right patient at the right time,  optimal dynamic treatment regimes, pioneered by \cite{murphy2003optimal} and \cite{robins2004optimal} have gained more and more attention in recent years \citep{chakraborty2013statistical,kosorok2015adaptive,tsiatis2019dynamic,laber2024handbook}. Dynamic treatment regimes (DTRs), in contrast to the population average treatment effect (ATE), highlight (i) the heterogeneity across individuals in treatment effect based on the fact that a treatment may have a negligible effect when averaged over all patients but could be beneficial for certain patient subgroups \citep{powers2018some} and (ii) the heterogeneity in treatment effect across time within an individual in response to their evolving characteristics, which are especially important in chronic diseases. An optimal DTR is the one leading to the greatest benefit (defined on the basis of pertinent outcomes) on average if followed by the entire population of interest. Methods for estimating optimal DTRs can be generally classified into indirect and direct approaches based on data from either randomized experiments or observational studies \citep{laber2014dynamic,barto2004reinforcement}. Basically, an indirect method models in a parametric, semiparametric, or non-parametric manner the conditional distribution of the outcome (or updated outcome) from which the optimal treatment rule can be straightforwardly derived for each decision point. Q-learning \citep{schulte2014q,laber2014interactive}, A-learning \citep{robins2004optimal,murphy2003optimal} and causal tree \citep{athey2016recursive} are standard and commonly used indirect methods developed based on different modelling assumptions. A direct method directly works on the defined average benefit and searches for the optimal regime which maximizes the objective over a pre-specified class of candidate regimes. Augmented inverse probability weighted estimator \citep{zhang2012robust,zhang2013robust} and outcome weighted learning \citep{zhao2012estimating,zhao2015new} are examples of direct methods. 

We restrict this tutorial to parsimonious and interpretable statistical models to learn optimal DTRs from observed data. In fact, lots of more flexible methods have been proposed, including non-parametric Q-learning \citep{taylor2015reader,zhou2017causal,qian2011performance}, V-learning \citep{luckett2020estimating} and super-learning \citep{luedtke2016super,williams2024learning}. These methods are powerful in capturing highly non-linear profiles and making accurate prediction, and have demonstrated their value in applications such as mobile health \citep{,deliu2024reinforcement}. However, they usually lead to complicated treatment decision rules with limited interpretability \citep{zhang2018interpretable,sun2021stochastic}.  To tackle the tension between model interpretability and prediction accuracy, rule-based learning methods have been developed, which yield easy-to-interpret decision rules, such as tree-based rules \citep{laber2015tree,tao2018tree} and list-based rules \citep{zhang2018interpretable} while employing flexible working models.  

This tutorial aims to provide a self-contained and accessible introduction to optimal DTRs. For readers who are interested in a more comprehensive review of this topic, we  recommend \cite{kosorok2019precision}, \cite{li2023optimal} and \cite{chakraborty2014dynamic} for further reading. The remainder of this tutorial is organized as follows. Section \ref{sec:assump} presents the definition and formulation of this topic in the casual inference framework and the identification assumptions required to link the causal quantity to the observed data. Section \ref{sec:model} introduces standard statistical models developed to estimate the optimal regime from data. In Section \ref{sec:sim}, we conduct simulation studies under two settings to illustrate the implementation and performance of different methods thereby providing further guidance. A real data application is given in Section \ref{sec:app}, followed by a discussion in Section \ref{sec:discussion}. R code is provided in the supplementary R markdown file \url{https://github.com/cywang0315/R_code_estimating_optimal_DTRs}. 

\section{Notations and framework} \label{sec:assump}
Suppose we have observed data collected from $n$ individuals in a longitudinal follow-up study. For each individual $i$, the corresponding records are collected in time order as $O_i=(L_{1i},A_{1i},\dots,L_{Ki},A_{Ki},Y_i), i=1,\dots,n$, where $A_{ji}$ is the observed treatment assignment, taking values in $\mathcal{A}_{j}$, a finite set of all possible treatment options with elements $a_j$ at the $j$th decision time; $L_{ji}$ is the covariate information (a vector of intermediate outcomes) collected subsequent to $A_{j-1,i}$ but prior to $A_{ji}$; and $Y_i$ is the observed final outcome of which, without loss of generality, larger values indicate better reward. We assume that the decision time points, indexed by $j$, for $j=1,\dots,K$ are common across individuals and that no death or dropout occurs during the follow-up period. Moreover, $\{O_i\}_{i=1}^{n}$ are assumed to be independent and identically distributed (i.i.d.) and the subscript $i$ may be omitted subsequently for simplicity and ease of exposition. To formalize an optimal dynamic treatment regime, we introduce the counterfactual framework. For a given treatment history $\bar{a}:=\bar{a}_{K}=(a_1,\dots,a_{K})$, we define $Y^{\ast}(\bar{a}
)$ as the potential final outcome that would have been observed if, possibly contrary to fact, the individual had followed treatment history $\bar{a}$. Similarly, $L^{\ast}_{j}(\bar{a}_{j-1})$, is the vector of potential intermediate outcomes that would have occurred if previously the individual had received treatment history $\bar{a}_{j-1}=(a_1,\dots,a_{j-1})$ for $j=2,\dots,K$. For the baseline covariates, $L_1^{\ast}$ is exactly the same as $L_1$ since no prior treatments are included. We then define $O^{\ast}$ to be the collection of all potential outcomes, i.e., $O^{\ast}=\{L_1^{\ast},\dots,L^{\ast}_{j}(\bar{a}_{j-1}),\dots,L^{\ast}_{K}(\bar{a}_{K-1}),Y^{\ast}(\bar{a}
) \text{ for all } \bar{a} \in \bar{\mathcal{A}}\}$, where $\bar{\mathcal{A}}:=\bar{\mathcal{A}}_{K}=\mathcal{A}_1 \times \dots \times \mathcal{A}_{K}$ is the set of all possible treatment histories $\bar{a}$ across all $K$ decisions. In general, for any given vector (random or not) $v$, we use the overbar to denote the prior (including the current) elements and use the underbar to denote the future (including the current) elements, that is, $\bar{v}_j=(v_1,\dots,v_j)$ and $\underline{v}_j=(v_j,\dots,v_K)$.

Before defining optimal dynamic treatment regimes, we firstly define dynamic treatment regimes. A dynamic treatment regime $d=(d_1,\dots,d_K)$ is a sequence of decision rules, one per decision point, indicating how treatment actions will be tailored over time to an individual's changing status \citep{murphy2003optimal}. Specifically, the $j$th rule $d_j:=d_j(\bar{l}_j,\bar{a}_{j-1})$, is a deterministic function, taking the covariate history $\bar{l}_j=(l_1,\dots, l_j)$ as well as the treatment history $\bar{a}_{j-1}$ prior to the $j$th decision time as inputs and outputting a specific treatment value $a_j \in \mathcal{A}_j$, $j=1,\dots,K$. In fact, the feasible set of output values usually depends on the input and is a subset of $\mathcal{A}_j$. For example, consider the case that once individuals start a particular treatment, they should remain on it thereafter. In this setting, the feasible output is always 1 if any element of input treatment history is 1. See Section 2 in \cite{schulte2014q} for a more rigorous statement on this issue.

Let $\mathcal{D}$ denote the class of well-defined dynamic treatment regimes. The optimal dynamic treatment regime is defined as the one that maximizes the expected potential final outcome over all eligible treatment regimes. That is,
\begin{equation}\label{1}
\begin{aligned}
d^{\text{opt}}&=\arg \max_{d\ \in \mathcal{D}} \mathbb{E} [Y^{\ast}(d)]\\
&= \arg \max_{d\ \in \mathcal{D}} \mathbb{E}\left[Y^{\ast}\left(a_1=d_1(L^{\ast}_1),\dots,a_K=d_K\left(\bar{L}^{\ast}_{K}(\bar{a}_{K-1}),\bar{a}_{K-1}\right)\right)\right]
\end{aligned}
\end{equation}
where $Y^{\ast}(d)$, defined by the second line in  (\ref{1}), denotes the potential outcome had regime $d$ been followed.
That is, we aim to find a particular treatment regime which will produce the most favourable final outcome on average if followed by the population. Note that the definition given by (\ref{1}) is applicable to a continuous final outcome $Y$. When the outcome of interest is an event time, an optimal dynamic treatment regime can be defined based on the expected log-transformed event time \citep{simoneau2020estimating} or the $\tau_0$-year survival probability with $\tau_0$ a predetermined time point \citep{jiang2017estimation}. 

The aforementioned definition of $d^{\text{opt}}$  does not easily facilitate its estimating procedure since (\ref{1}) involves all $K$ rules. Fortunately, dynamic programming transfers this problem into a sequence of simpler problems in a backward inductive manner as follows: starting from the $K$th stage,
\begin{equation}\label{2K}
\begin{aligned}
&d^{\text{opt}}_K(\bar{l}_{K},\bar{a}_{K-1})=\arg\max_{a_{K}\in \mathcal{A}_{K}} \mathbb{E}[Y^{\ast}(\bar{a}_{K-1},a_{K})|\bar{L}^{\ast}_{K}(\bar{a}_{K-1})=\bar{l}_K], \\
&V_{K}(\bar{l}_{K},\bar{a}_{K-1})=\max_{a_{K}\in \mathcal{A}_{K}} \mathbb{E}[Y^{\ast}(\bar{a}_{K-1},a_K)|\bar{L}^{\ast}_{K}(\bar{a}_{K-1})=\bar{l}_K];
\end{aligned}
\end{equation}
then for $j=K-1,\dots, 1$,
\begin{equation}\label{2j}
\begin{aligned}
&d^{\text{opt}}_{j}(\bar{l}_{j},\bar{a}_{j-1})=
\arg\max_{a_{j}\in \mathcal{A}_{j}}\mathbb{E}[V_{j+1}(\bar{l}_{j},L^{\ast}_{j+1}(\bar{a}_{j-1},a_{j}),\bar{a}_{j-1},a_{j})|\bar{L}^{\ast}_{j}(\bar{a}_{j-1})=\bar{l}_{j}],\\
&V_j(\bar{l}_j,\bar{a}_{j-1})=\max_{a_{j}\in \mathcal{A}_{j}} \mathbb{E}[V_{j+1}(\bar{l}_{j},L^{\ast}_{j+1}(\bar{a}_{j-1},a_{j}),\bar{a}_{j-1},a_{j})|\bar{L}^{\ast}_{j}(\bar{a}_{j-1})=\bar{l}_{j}].
\end{aligned}
\end{equation}
It can be shown that  $(d_1^{\text{opt}},\dots,d_K^{\text{opt}})$ given by (\ref{2K}) and (\ref{2j}) is an optimal dynamic treatment regime satisfying (\ref{1}) \citep{schulte2014q,robins2004optimal}.

Until now, $d^{\text{opt}}$ has been expressed in terms of potential outcomes $O^{\ast}$. To identify $d^{\text{opt}}$ from the observed data $\{O_{i}\}_{i=1}^{n}$, the following assumptions which are standard in the causal inference literature are required.
\begin{itemize}
\item (Consistency) $L_j^{\ast}(\bar{a}_{j-1})=L_j$ if $\bar{A}_{j-1}=\bar{a}_{j=1}$, for $j=1,\dots,K$; and $Y^{\ast}(\bar{a})=Y$ if $\bar{A}=\bar{a}$.
\item (Stable unit treatment value) The potential outcomes of any individual are unaffected by the
treatment assignment of any other individual.
\item (Positivity) If $P(\bar{L}_j=\bar{l}_j,\bar{A}_{j-1}=\bar{a}_{j-1})>0$, then $P(A_j=a_j|\bar{L}_j=\bar{l}_j,\bar{A}_{j-1}=\bar{a}_{j-1})>0$ for any $a_j \in \mathcal{A}_j$, $j=1,\dots,K$.
\item (Sequential randomization) $A_j \ind O^{\ast} |\bar{L}_j,\bar{A}_{j-1}$ for $j=1,\dots K$.
\end{itemize}

The consistency assumption links the potential outcomes to the observed data: the observed (actual) outcomes (intermediate and final) are exactly the potential outcomes under the treatments actually received \citep{robins1994correcting}. The stable unit treatment value assumption is also called the no-interference assumption \citep{rubin1980randomization}, which may be violated due to social interactions between individuals, for example, the effect of vaccination on an infectious disease in a cohort \citep{hudgens2008toward}. This tutorial is restricted to the case where the no-interference assumption holds. Please refer to \cite{su2019modelling} and \cite{jiang2023dynamic} for estimating optimal treatment regimes when interference is present. 

Positivity means that the treatment received at each stage $j$ was not deterministically allocated within any possible level $(\bar{l}_{j},\bar{a}_{j-1})$ of the history $(\bar{L}_{j},\bar{A}_{j-1})$; that is, each admissible treatment in stage $j$ has a positive probability of being chosen for all possible histories. If  $A_j$ is continuous, the probability should be replaced with the corresponding density function. Note that the positivity assumption above is necessary for non-parametric identification of the conditional causal effect of $A_j$ given $(\bar{l}_{j},\bar{a}_{j-1})$. When an unsaturated model for the treatment effect is specified, the positivity assumption can be relaxed and the identification can be achieved due to model-based extrapolation \citep{robins2008estimation}.

Sequential randomization is also called sequential ignorability, conditional exchangeability or no unmeasured confounders, which states that the treatment assignment at each stage is independent of the set of potential outcomes $O^{\ast}$, given the covariate and treatment histories.  There are several weaker versions of the sequential randomization assumption proposed in the causal inference context. For example, when static (non-dynamic) treatment regimes are of interest, the assumption required is $A_j \ind Y^{\ast}(\bar{a}_{j-1},\mathbf{0})|\bar{L}_j,\bar{A}_{j-1}=\bar{a}_{j-1}$ for any $\bar{a}_{j-1} \in \bar{\mathcal{A}}_{j-1}$ and $j=1,\dots K$ \citep{vansteelandt2014structural}. That is, $A_j$ is assumed to be independent of the final potential outcome when the future treatments are fixed to be zero (the reference level); whereas for the setting of dynamic treatment regimes considered here, potential outcomes (both intermediate and final) under various future treatments are included due to the dynamic nature concerned. This (strong) sequential randomization assumption is essential in establishing the equivalence between the conditional distribution of the potential outcomes and that of the observed outcomes, i.e., the distribution of $Y^{\ast}(\bar{a}_{K})|\bar{L}^{\ast}_{K}(\bar{a}_{K-1})$ (or $L_j^{\ast}(\bar{a}_{j-1})|\bar{L}_{j-1}^{\ast}(\bar{a}_{j-2})$ for $j=1,\dots,K$) versus the distribution of $Y|\bar{L}_{K},\bar{A}_{K}=\bar{a}_{K}$ (or $L_j|\bar{L}_{j-1}, \bar{A}_{j-1}=\bar{a}_{j-1}$ for $j=1,\dots,K$). See Section A.2 in the supplement material of \cite{vansteelandt2014structural} and Lemma 3.1 in \cite{robins2004optimal} for a detailed demonstration.  See \cite{cui2021semiparametric} and \cite{pu2021estimating} for identifying optimal treatment regimes (in the case of a single stage) in the presence of unmeasured confounding.  

Under assumptions specified above, the optimal dynamic treatment regime defined by (\ref{2K}) and (\ref{2j}) can be re-expressed in terms of the observed data. Let $Q_K(\bar{l}_K,\bar{a}_{K})=\mathbb{E}[Y|\bar{L}_{K}=\bar{l}_K,\bar{A}_{K}=\bar{a}_{K}]$, then $d_{K}^{\text{opt}}$ in (\ref{2K}) becomes
\begin{equation}\label{3}
\begin{aligned}
&d^{\text{opt}}_K(\bar{l}_{K},\bar{a}_{K-1})=\arg\max_{a_{K}\in \mathcal{A}_{K}} Q_K(\bar{l}_K,\bar{a}_{K-1},a_{K}),
\end{aligned}
\end{equation}
and $V_{K}(\bar{l}_{K},\bar{a}_{K-1})$ becomes
\begin{equation}\label{4}
\begin{aligned}
&V_{K}(\bar{l}_{K},\bar{a}_{K-1})=\max_{a_{K}\in \mathcal{A}_{K}}Q_K(\bar{l}_K,\bar{a}_{K-1},a_{K}).
\end{aligned}
\end{equation}
Similarly, for $j=K-1,\dots1$, let 
\begin{equation}\label{5}
Q_j(\bar{l}_j,\bar{a}_{j})=\mathbb{E}[V_{j+1}(\bar{l}_{j},L_{j+1},\bar{a}_{j})|\bar{L}_{j}=\bar{l}_{j},\bar{A}_{j}=\bar{a}_{j}].
\end{equation}
Then for $d^{\text{opt}}_{j}(\bar{l}_{j},\bar{a}_{j-1})$ in (\ref{2j}), we have
\begin{equation}\label{6}
\begin{aligned}
&d^{\text{opt}}_{j}(\bar{l}_{j},\bar{a}_{j-1})=
\arg\max_{a_{j}\in \mathcal{A}_{j}} Q_j(\bar{l}_j,\bar{a}_{j-1},a_j),
\end{aligned}
\end{equation}
and
\begin{equation}\label{7}
\begin{aligned}
&V_j(\bar{l}_j,\bar{a}_{j-1})=\max_{a_{j}\in \mathcal{A}_{j}} Q_j(\bar{l}_j,\bar{a}_{j-1},a_j).
\end{aligned}
\end{equation}
Usually, $Q_j(\bar{l}_j,\bar{a}_{j})$ is referred to as the Q-function, which measures the quality of assigning treatment $a_j$ to patients given the history $(\bar{l}_j,\bar{a}_{j-1})$ and with the optimal regime followed thereafter; and $V_j(\bar{l}_j,\bar{a}_{j-1})$ (also denoted by $V_j$ for brevity) is called the value function, which is the maximum  value of $Q_j(\bar{l}_j,\bar{a}_{j})$ achieved at $a_j=d^{\text{opt}}_{j}(\bar{l}_{j},\bar{a}_{j-1})$ given the history $(\bar{l}_j,\bar{a}_{j-1})$ for $j=1,\dots,K$. Moreover, it can be shown that 
$$V_j(\bar{l}_j,\bar{a}_{j-1})=\mathbb{E}[Y^{\ast}(\bar{a}_{j-1},\underline{d}^{\text{opt}}_{j})|\bar{L}^{\ast}_{j}(\bar{a}_{j-1})=\bar{l}_{j}]=\mathbb{E}[Y^{\ast}(\bar{a}_{j-1},\underline{d}^{\text{opt}}_{j})|\bar{L}_{j}=\bar{l}_{j},\bar{A}_{j-1}=\bar{a}_{j-1}]$$ 
where the second equality holds due to the sequential randomization assumption. In particular, $\mathbb{E}[V_1(L_1)]=\mathbb{E}[Y^{\ast}(d^{\text{opt}})]$. The procedure given by (\ref{3})-(\ref{7}) allows us to learn the optimal dynamic treatment regimes from the observed data by introducing statistical models.

\section{Statistical approaches} \label{sec:model}
In this section, we introduce some commonly used statistical methods to estimate optimal DTRs from observed data and restrict our attention to the case where the treatment variable at each stage is binary.  We use a simple simulated example throughout the section to illustrate how a specific method works. This example, as shown in Table \ref{tab:sim_data} for three individuals, involves two decision points $K=2$ with one-dimensional covariate information collected prior to each treatment action (i.e. $L_1$ prior to $A_1$ and $L_2$ prior to $A_2$) and a final outcome $Y$. Section \ref{sec:simcase1} describes the data-generating model behind this example.
\begin{table}[h]
\centering
\caption{\label{tab:sim_data}  Follow-up records for three random individuals ($n=500$) in the simulated example.}
\begin{tabular*}{0.57\linewidth}{cccccc}
 \hline
Individual &$L_1$ & $A_1$ & $L_2$ & $A_2$ & $Y$\\   \hline
1&356.03 & 0 &357.76&  0& 1016.17\\
2&477.55  &0 &546.19&  0& 1108.56\\
3&324.65  &1 &330.79&  1&  898.41\\
  \hline
\end{tabular*}
\end{table}

\subsection{Indirect methods}\label{sec:indirect methods}
Indirect methods are usually regression-based approaches, which completely (e.g., Q-learning) or partially  (e.g., A-learning) specify how the (updated) final outcome depends on the covariate and treatment histories. From the estimated regression model, the optimal treatment rule can be derived in a straightforward way. Here we focus on Q-learning, A-learning and causal tree (which can be viewed as a non-parametric A-learning method). We emphasize the different modelling assumptions behind them and outline the unified backward induction algorithm for their implementation.

\subsubsection{Q-learning}\label{sec:Q-learning}
Q-learning is developed directly from (\ref{3})-(\ref{7}) by modelling the Q-functions (conditional mean functions)  in  (\ref{5}) and then deriving the optimal regimes based on the estimated conditional means. Specifically, a parametric model $Q_j(\bar{l}_{j},\bar{a}_j;\theta_j), j=1,\dots,K$ is specified in each stage with a finite-dimensional parameter $\theta_j$ representing the possible main and interaction effects of $\bar{l}_j$ and $\bar{a}_j$ on the value. Note that $a_j$ in $Q_j(\bar{l}_{j},\bar{a}_j;\theta_j)$ should be given special attention since the next step involves optimizing over $a_j$. To make this explicit, we separate out $a_j$. Let $H_j$ denote the covariate and treatment history available before the $j$th decision time, i.e., $H_j=(L_1,A_1,\dots,L_{j-1},A_{j-1},L_j)$ and $h_j$ be a realization of $H_j$ for $j=1,\dots, K$. Then the following additive model can be used to specify Q-functions:  
\begin{equation}\label{Q-full}
\begin{aligned}
Q_j(\bar{L}_{j},\bar{A}_j;\theta_j)=Q_j(H_{j},A_j;\psi_j,\xi_j)=A_jC_j(H_j;\psi_j)+m_j(H_j;\xi_j),\quad j=1,\dots,K,
\end{aligned}
\end{equation}
where $C_j(H_j;\psi_j)$ is called the Q-contrast function (or 
contrast function for brevity), characterizing the dynamic nature of the treatment, i.e., how the effect of $a_j$ varies in response to different histories; $m_j(H_j;\xi_j)=Q_j(H_{j},0;\psi_j,\xi_j)$ is the treatment-free (free of $a_j$) term, which specifies the profile of the outcome when the $j$th treatment is set to the reference treatment defined as 0 in the binary treatment scenario; and $\theta_j=(\psi_j,\xi_j)$ includes parameters in the two terms with $\psi_j$ and $\xi_j$ parameterizing the contrast function and the treatment-free term, respectively. Moreover, the parameterization of the contrast function should satisfy $C_j(h_j;0)=0$ for all $h_j$ so that $\psi_j=0$ indicates the null hypothesis of no treatment effect of $A_j$ when optimal regimes followed thereafter. Once an estimate of $\theta_j$, say, $\hat{\theta}_{j}$ is obtained, it is straightforward to derive the corresponding estimate of $d_j^{\text{opt}}$: $\hat{d}_j^{\text{opt}}:=\hat{d}_j^{\text{opt}}(H_j)=I\left\{C_j(H_j;\hat{\psi}_j)>0\right\}$, and to construct the next (i.e., the $(j-1)$th stage) response variable by substituting $\hat{d}_j^{\text{opt}}$ for $a_j$ in (\ref{Q-full}): $\widehat{V}_{j}^{Q}:=V_{j}(\bar{L}_{j},\bar{A}_{j-1};\hat{\theta}_{j})=I\left\{C_j(H_j;\hat{\psi}_j)>0\right\}C_j(H_j;\hat{\psi}_j)+m_j(H_j;\hat{\xi}_j)$.

In our simulated example, we specify the Q-function in stage 2 as follows:
\begin{equation}
Q_2(H_2,A_2;\psi_2,\xi_2)=A_2R_2^{\top}\psi_2+D_2^{\top}\xi_2, \label{Q2_example}
\end{equation}
where $R_2=(1,L_2)^{\top}$ and $D_2=(1,L_1,A_1,L_1A_1,L_2)^{\top}$. Fitting (\ref{Q2_example}) to the simulated dataset (of sample size 500) using ordinary least squares (OLS) with $Y$ being the response yields $\hat{\psi}_2=(479.88, -1.54)^{\top}$ and $\hat{\xi}_2=(284.08, 1.55, 43.81, -0.53, 0.21)^{\top}$, which gives rise to  $\hat{d}_2^{\text{opt}}=I\left\{479.88-1.54L_2>0\right\}$, i.e., treat if $L_2<479.88/1.54=312$ (see Section \ref{sec:simcase1} for the true $d_2^{\text{opt}}$). As mentioned in the previous paragraph, the estimated value function would now be $\widehat{V}_{2}^{Q}=I\left\{479.88-1.54L_2>0\right\}(479.88-1.54L_2)+D_2^{\top}\hat{\xi}_2$, and will be used as the response variable in stage 1. As can be seen in Table \ref{tab:sim_data_Q_stage2},  $\widehat{V}_{2}^{Q}$ might be smaller than $Y$ for some individuals due to (i) the magnitude of the conditional variance of $Y$ given the history and (ii) the misspecification of $Q_2$. 
In stage 1, the Q-function is specified as 
\begin{equation}
Q_1(H_1,A_1;\psi_1,\xi_1)=A_1R_1^{\top}\psi_1+D_1^{\top}\xi_1, \label{Q1_example}
\end{equation}
where $R_1=D_1=(1,L_1)^{\top}$, and the estimation of $(\psi_1,\xi_1)$ is performed based on the variables highlighted in blue in Table \ref{tab:sim_data_Q_stage2}. We get $\hat{\psi}_1=(167.13, 
 -0.82)^{\top}$ and $\hat{\xi}_1=(337.14, 1.71)^{\top}$ from OLS, and therefore  $\hat{d}_1^{\text{opt}}=I\left\{167.13-0.82L_1>0\right\}$, i.e., treat if $L_1<167.13/0.82=204$ and $\widehat{V}_{1}^{Q}=I\left\{167.13-0.82L_1>0\right\}(167.13-0.82L_1)+D_1^{\top}\hat{\xi}_1$. The average values (over 500 individuals) of $\widehat{V}_{1}^{Q}$, $\widehat{V}_{2}^{Q}$ and $Y$ are  $1114.19$, $1059.36$ and $999.01$, respectively.
\begin{table}[h]
\centering
\caption{\label{tab:sim_data_Q_stage2}  The predicted optimal treatment actions and value functions from Q-learning for the three random individuals in the simulated example.}
\begin{tabular*}{0.92\linewidth}{@{\extracolsep{\fill}} cccccccccc}
\hline \vspace{-0.3cm}\\
Individual & \textcolor{blue}{$L_1$} &\textcolor{blue}{$A_1$} & $L_2$ & $A_2$ & $Y$ & $\hat{d}_2^{\text{opt}}$ & \textcolor{blue}{$\widehat{V}_{2}^{Q}$} &  $\hat{d}_1^{\text{opt}}$ & $\widehat{V}_{1}^{Q}$ \\   \hline
1& \textcolor{blue}{356.03} & \textcolor{blue}{0} &357.76&  0& 1016.17 & 0 & \textcolor{blue}{908.05} & 0 & 945.95\\
2& \textcolor{blue}{477.55} &\textcolor{blue}{0} &546.19&  0& 1108.56 &0 & \textcolor{blue}{1134.66} & 0 & 1153.74 \\
3&\textcolor{blue}{324.65} &\textcolor{blue}{1} &330.79&  1&  898.41 &0 & \textcolor{blue}{724.67} &0 & 892.30\\
  \hline
\end{tabular*}
\end{table}

In summary,  Q-learning proceeds in a backward iterative way: staring from the $K$th stage, we obtain $\hat{\theta}_K$ by fitting $Q_K(\bar{l}_{K},\bar{a}_K;\theta_K)$ via ordinary least squares (OLS) with $Y$ as the response and derive $\hat{d}^{\text{opt}}_K$ and $\tilde{V}_{K}^{Q}$ based on $\hat{\theta}_K$; then we move one step backwards to the $(K-1)$th stage and fit $Q_{K-1}(\bar{l}_{K-1},\bar{a}_{K-1};\theta_{K-1})$ to get $\hat{\theta}_{K-1}$ with  $\tilde{V}_{K}^{Q}$ now as the response; and continue this process until $\hat{d}_1^{\text{opt}}$ is obtained. Q-learning is clear and easy to understand because its procedure exactly follows the definition of $d^{\text{opt}}$ in (\ref{3})-(\ref{7}). The performance of Q-learning heavily depends on the specification of the Q-functions. The estimated regime $\hat{d}^{\text{opt}}=(\hat{d}_1^{\text{opt}},\dots,\hat{d}_K^{\text{opt}})$ may not be a consistent estimate of the true $d^{\text{opt}}$ unless all the models for Q-functions are correctly specified. However, in some cases, it is impossible to correctly specify all the Q-functions using standard regression models due to the dependence between any two adjacent Q-functions implied by their definitions. To illustrate, consider the case that $Q_K(\bar{l}_{K},\bar{a}_K;\theta_K)$ is specified as a linear regression model, i.e., $Q_K(\bar{l}_{K},\bar{a}_K;\theta_K)=a_Kr_K^{\top}\psi_K+s^{\top}_{K}\xi_K$, where $r_K=s_K=(1,l_{K}^{\top})^{\top}$ and that $L_j$ in each stage is a scalar covariate. Then  the correct expression of $Q_{K-1}(\bar{l}_{K-1},\bar{a}_{K-1})$ can be rigorously derived:
\begin{equation}\label{Q-incompata}
\begin{aligned}
Q_{K-1}(\bar{l}_{K-1},\bar{a}_{K-1})=\mathbb{E}\Big[&I\left\{\psi_{K0}+\psi_{K1}L_{K}>0\right\}(\psi_{K0}+\psi_{K1}L_{K})\\
&+(\xi_{K0}+\xi_{K1}L_{K})|\bar{L}_{K-1}=\bar{l}_{K-1},\bar{A}_{K-1}=\bar{a}_{K-1}\Big],
\end{aligned}
\end{equation}
where the expectation is with respect to $L_{K}|\bar{L}_{K-1}=\bar{l}_{K-1},\bar{A}_{K-1}=\bar{a}_{K-1}$. Obviously, the indicator function involved in the first term in (\ref{Q-incompata}) makes $Q_{K-1}(\bar{l}_{K-1},\bar{a}_{K-1})$ a highly nonlinear function of $\bar{l}_{K-1}$, which will be poorly approximated by a linear regression model. See Section 5.3 in \cite{schulte2014q} for a derived expression of (\ref{Q-incompata}) when $L_{K}|\bar{L}_{K-1},\bar{A}_{K-1}$ follows a normal distribution.

To address the misspecification issue in Q-learning, \cite{laber2014interactive} proposed interactive Q-leaning (IQ-learning), which attempts to estimate the two terms on the RHS of (\ref{Q-incompata}) separately: 
\begin{equation*}
\begin{aligned}
\widehat{Q}_{K-1}^{\text{IQ}}(\bar{l}_{K-1},\bar{a}_{K-1})=\int_{0}^{\infty} x \hat{p}_{\bar{l}_{K-1},\bar{a}_{K-1}}(x)dx+\widehat{B}(\bar{l}_{K-1},\bar{a}_{K-1}),
\end{aligned}
\end{equation*}
where $\hat{p}_{\bar{l}_{K-1},\bar{a}_{K-1}}(\cdot)$ is an estimator of the density function of $\psi_{K0}+\psi_{K1}L_{K}$ (i.e., $C_K(H_K;\psi_K)$) conditional on $\bar{L}_{K-1}=\bar{l}_{K-1}$ and $\bar{A}_{K-1}=\bar{a}_{K-1}$; and $\widehat{B}(\bar{l}_{K-1},\bar{a}_{K-1})$ is an estimator of $\mathbb{E}[\xi_{K0}+\xi_{K1}L_{K}|\bar{L}_{K-1}=\bar{l}_{K-1},\bar{A}_{K-1}=\bar{a}_{K-1}]$ for which linear models are usually adequate. As for  $\hat{p}_{\bar{l}_{K-1},\bar{a}_{K-1}}$, mean-variance function modelling approach is used, which captures the dependence of $C_K(H_K;\psi_K)$ on $(H_{K-1},A_{K-1})$ via conditional mean and variance functions. Besides IQ-learning, extending the standard regression models for Q-functions to a more flexible modelling framework has been extensively investigated, such as generalized additive models \citep{moodie2014q}, support vector regression \citep{zhao2009reinforcement,zhao2011reinforcement} and kernel ridge regression \citep{zhang2018interpretable}.

\subsubsection{A-learning}
In contrast to Q-learning which specifies the full Q-functions, advantage learning (A-learning) focuses on the modelling of contrast functions, or equivalently, regret functions. A-learning can be understood from different perspectives. On the one hand, A-learning attempts to model and estimate the Q-contrast function $C_j(h_j;\psi_j)$ without any specification on $m_j(h_j)$ as this is all that is needed to make a decision. That is, instead of specifying a fully parametric modelling form as in (\ref{Q-full}), we generalise to the semi-parametric model:
\begin{equation}\label{Q-semi}
\begin{aligned}
Q_j(H_{j},A_j;\psi_j)=A_jC_j(H_j;\psi_j)+m_j(H_j),\quad j=1,\dots,K,
\end{aligned}
\end{equation}
where $C_j(\cdot)$ is specified parametrically but $m_j(\cdot)$ is left unspecified and treated non-parametrically. For estimation, $\psi_K$ can be estimated by \citep{robins1992estimating}
\begin{equation}\label{EE_K}
\begin{aligned}
\sum_{i=1}^{n}\frac{\partial C_{K}(H_{Ki};\psi_K)}{\partial \psi_K}\{Y_i-A_{Ki}C_{K}(H_{Ki};\psi_K)\}\{A_{Ki}-\mathbb{E}[A_{Ki}|H_{Ki}]\}=0,
\end{aligned}
\end{equation}
which is an unbiased estimating equation (EE) for $\psi_K$ based on the following fact implied by (\ref{Q-semi}):
$$\mathbb{E}[Z_j|H_j,A_j]=\mathbb{E}[Z_j|H_{j}],\quad j=1,\dots,K,$$ 
with $Z_K=Y-A_KC_K(H_K;\psi_K)$ and $Z_j=V_{j+1}(\bar{L}_{j+1},\bar{A}_j)-A_jC_j(H_j;\psi_j)$ for $j=1,\dots,K-1$. Given an estimate of $\psi_j$ (and also $\psi_{j+1},\dots,\psi_{K}$), the optimal decision rule for the $j$th stage is $\hat{d}^{\text{opt}}_j=I\left\{C_j(H_j;\hat{\psi}_j)>0\right\}$; and the response variable for stage $j-1$ is constructed by 
\begin{equation}\label{response_A}
\begin{aligned}
\widehat{V}_{j}^{A}=Y+\sum_{k=j}^{K}\left(I\left\{C_k(H_k;\hat{\psi}_k)>0\right\}-A_k\right)C_k(H_k;\hat{\psi}_k),
\end{aligned}
\end{equation}
which satisfies $\mathbb{E}[\widehat{V}_{j}^{A}|H_j,A_j]=V_j(\bar{L}_j,\bar{A}_{j-1})$ for $j=1,\dots,K$. Then $\psi_{j-1}$ can be estimated by solving an estimating equation similar to (\ref{EE_K}) with $Y_i$ replaced by $\widehat{V}_{j}^{A}$ for $j=1,\dots,K$. As we can see, A-learning also proceeds backward through stages. It differs from Q-learning in (i) an explicit model is specified for the contrast functions only; (ii) an estimating equation, e.g., (\ref{EE_K}), developed for the semiparametric model is used in each stage; (iii) a  model for the propensity score $\mathbb{E}[A_j|H_j]$ is required; and (iv) the response variable for the next stage is constructed by (\ref{response_A}) to ensure its expectation equals $V_j$ while in Q-learning $V_j$ itself is explicitly derived from Q-functions.

On the other hand, contrast-based A-learning, as pointed out in \cite{robins2004optimal}, is closely related to structural nested mean models (SNMMs), which were initially proposed to estimate the effect of a time-dependent treatment in the presence of confounders. An SNMM models the contrast between $Y^{\ast}(\bar{a}_{j},\underline{0}_{j+1})$ and $Y^{\ast}(\bar{a}_{j-1},\underline{0}_{j})$, conditional on the past covariate and treatment history:
\begin{equation}\label{SNMM}
\begin{aligned}
\mathbb{E}[Y^{\ast}(\bar{a}_{j},\underline{0}_{j+1})-Y^{\ast}(\bar{a}_{j-1},\underline{0}_{j})|H_j=h_j,A_j=a_j]=\gamma_j^{\text{ref}}(h_j,a_j;\psi), \quad j=1,\dots,K,
\end{aligned}
\end{equation}
where $\gamma_j^{\text{ref}}(h_j,a_j;\psi)$ is called the blip-to-reference function since it measures the effect of a blip of treatment $a_j$ versus zero (reference) treatment at stage $j$ with all future treatments fixed at their reference level 0. Obviously, a proper parametrization of the underlying function $\gamma_j^{\text{ref}}(h_j,a_j)$ should satisfy: $\gamma_j^{\text{ref}}(h_j,0;\psi)=0$ and $\gamma_j^{\text{ref}}(h_j,a_j;0)=0$. Let $U_j^{\text{ref}}(\psi)=Y-\sum_{k=j}^{K}\gamma_k^{\text{ref}}(H_k,A_k;\psi)$, which removes the effect of the observed treatments over stages $j$ to $K$. 
Then from model (\ref{SNMM}) and the consistency assumption, we have 
\begin{equation*}
\begin{aligned}
\mathbb{E}[U_j^{\text{ref}}(\psi)|\bar{L}_j,\bar{A}_{j}=\bar{a}_{j}]=\mathbb{E}[Y^{\ast}(\bar{a}_{j-1},\underline{0}_{j})|\bar{L}_j,\bar{A}_{j}=\bar{a}_{j}],
\end{aligned}
\end{equation*}
which, together with a weaker exchangeability assumption $A_j \ind Y^{\ast}(\bar{a}_{j-1},\underline{0}_{j})|\bar{L}_j,\bar{A}_{j-1}=\bar{a}_{j-1}$, yields
\begin{equation}\label{con_indep}
\begin{aligned}
\mathbb{E}[U_j^{\text{ref}}(\psi)|H_j,A_{j}]=\mathbb{E}[U_j^{\text{ref}}(\psi)|H_j],\quad j=1,\dots,K.
\end{aligned}
\end{equation}
Estimation of $\psi$ can thus be performed by solving
\begin{equation}\label{EE_SNMM}
\begin{aligned}
\sum_{i=1}^{n}\sum_{j=1}^{K}\left\{U_{j,i}^{\text{ref}}(\psi)-\mathbb{E}\left[U_{j,i}^{\text{ref}}(\psi)|H_{ji}\right]\right\}\left\{g_j(H_{ji},A_{ji})-\mathbb{E}\left[g_j(H_{ji},A_{ji})|H_{ji}\right]\right\}=0,
\end{aligned}
\end{equation}
which sets the empirical conditional covariance between $U_{j}^{\text{ref}}(\psi)$ and  $g_j(H_j,A_j)$ (an arbitrary vector function of the same dimension as $\psi$) given $H_j$, to zero \citep{vansteelandt2014structural}, and is referred to as g-estimation of SNMMs.  Working models for $\mathbb{E}\left[U_{j}^{\text{ref}}(\psi)|H_{j}\right]$ and $\mathbb{E}\left[g_j(H_{j},A_{j})|H_{j}\right]$ are required to solve (\ref{EE_SNMM}) and the estimate of $\psi$ from  (\ref{EE_SNMM}) is consistent when one of the working models is correctly specified. See \cite{comments} for discussions on double-robust estimators.

To investigate optimal DTRs, \cite{robins2004optimal} extended the SNMM in (\ref{SNMM}) to the optimal double-regime structural nested mean model (drSNMM):
\begin{equation}\label{opt_SNMM}
\begin{aligned}
\mathbb{E}[Y^{\ast}(\bar{a}_{j},\underline{d}_{j+1}^{\text{opt}})-Y^{\ast}(\bar{a}_{j-1},0,\underline{d}_{j+1}^{\text{opt}})|H_j=h_j,A_j=a_j]=\gamma_j(h_j,a_j;\psi_j), \quad j=1,\dots,K,
\end{aligned}
\end{equation}
which models the effect of a
blip of treatment $a_j$ versus zero (reference) treatment at stage $j$ when the optimal treatment regime is followed from $j+1$ onwards, and is also called the optimal blip function in \cite{moodie2007demystifying}. Note that the optimal drSNMM in (\ref{opt_SNMM}) differs from SNMM in: (i) double regimes are involved, one is the reference treatment in stage $j$ and the other one is the optimal dynamic regime followed in the future and (ii) common \textit{treatment rules} $\underline{d}_{j+1}^{\text{opt}}$ rather than common \textit{treatment values} are followed subsequently. In fact, the actual treatment of individuals following $(\bar{a}_{j},\underline{d}_{j+1}^{\text{opt}})$ may differ from those following $(\bar{a}_{j-1},0,\underline{d}_{j+1}^{\text{opt}})$ at times subsequent to stage $j$ (see remark 3.2 in \cite{robins2004optimal}). For estimation of $\psi_j$, the estimating equation in the $j$th stage can be constructed based on a similar property to (\ref{con_indep}). Specifically, let 
\begin{equation}\label{counter_outcome}
\begin{aligned}
U_j(\psi_j)=Y-\gamma_j(H_j,A_j;\psi_{j})+\sum_{k=j+1}^{K}\left[\gamma_k(H_k,\hat{d}_k^{\text{opt}};\hat{\psi}_{k})-\gamma_k(H_k,A_k;\hat{\psi}_{k})\right],
\end{aligned}
\end{equation}
where $\hat{\psi}_{j+1},\dots,\hat{\psi}_{K}$ are estimates of parameters over stages $j+1$ to $K$ and are assumed to have already been obtained when estimating $\psi_j$; $\hat{d}_k^{\text{opt}}, k=j+1,\dots,K$ are the corresponding estimated optimal treatment rules $\hat{d}_k^{\text{opt}}=\arg\max_{a_k} \gamma_k(h_k,a_k;\hat{\psi}_k)$ for $k=j+1,\dots,K$. Note that $U_j(\psi_j)$ intuitively removes the effect of the observed treatment $A_j$ from $Y$ by subtracting $\gamma_j(H_j,A_j;\psi_{j})$, and also replaces the subsequent observed treatments with the treatments specified by the optimal decision rules by adding back $\gamma_k(H_k,\hat{d}_k^{\text{opt}};\hat{\psi}_{k})-\gamma_k(H_k,A_k;\hat{\psi}_{k})$ for $k=j+1,\dots,K$. In fact, if the optimal blip functions are correctly specified and $\hat{\psi}_k$ for $k=j+1,\dots,K$ are unbiased estimates, we have 
\begin{equation*}
\begin{aligned}
\mathbb{E}[U_j(\psi_j)|\bar{L}_j,\bar{A}_{j}=\bar{a}_{j}]=\mathbb{E}[Y^{\ast}(\bar{a}_{j-1},0,\underline{d}^{\text{opt}}_{j+1})|\bar{L}_j,\bar{A}_{j}=\bar{a}_{j}],
\end{aligned}
\end{equation*}
and further
\begin{equation*}
\begin{aligned}
\mathbb{E}[U_j(\psi_j)|H_j,A_{j}]=\mathbb{E}[U_j(\psi_j)|H_j],\quad j=1,\dots,K,
\end{aligned}
\end{equation*}
under assumptions in Section \ref{sec:assump}. Therefore, $\psi_j$ can be estimated by solving 
\begin{equation}\label{EE_optSNMM}
\begin{aligned}
\sum_{i=1}^{n}\left\{U_{j,i}(\psi_j)-\mathbb{E}\left[U_{j,i}(\psi_j)|H_{ji}\right]\right\}\left\{g_j(H_{ji},A_{ji})-\mathbb{E}\left[g_j(H_{ji},A_{ji})|H_{ji}\right]\right\}=0,
\end{aligned}
\end{equation}
where $g_j(H_j,A_j)$ is an arbitrary function of the dimension of $\psi_j$. One choice of $g_j(H_j,A_j)$ is to set 
$g_j(H_j,A_j)=\mathbb{E}\left[\partial U_j(\psi_j)/\partial \psi_j |H_j,A_j\right].$
Different choices of $g_j(H_j,A_j)$ will make an impact on the efficiency of the parameter estimates in (\ref{EE_optSNMM}) (also (\ref{EE_SNMM})). See Section 3.3 in \cite{robins2004optimal} for discussions on efficiency. Moreover, the inclusion of both $\mathbb{E}\left[S_{j}(\psi_j)|H_{j}\right]$ (equivalently, $\mathbb{E}[Y^{\ast}(\bar{A}_{j-1},0,\underline{d}^{\text{opt}}_{j+1})|H_j]$) and $\mathbb{E}\left[g_j(H_{j},A_{j})|H_{j}\right]$ achieves the double protection against model misspecification. Note that the model for $\mathbb{E}\left[g_j(H_{j},A_{j})|H_{j}\right]$ can be fitted separately. However, the model for  $\mathbb{E}\left[U_{j}(\psi_j)|H_{j}\right]$ has to be fitted jointly with (\ref{EE_optSNMM}) since the response variable $U_{j}(\psi_j)$ depends on the unknown $\psi_j$. A closed-form of $\hat{\psi}_{j}$ from (\ref{EE_optSNMM}) can be obtained when the optimal blip function is linear in $\psi_j$ and concurrently a linear regression model is specified for $\mathbb{E}\left[U_{j}(\psi_j)|H_{j}\right]$. That's also the reason why $U_j(\psi_j)$ is defined by (\ref{counter_outcome}) rather than
\begin{equation*}
\begin{aligned}
Y+\gamma_j(H_j,d^{\text{opt}}_j;\psi_{j})-\gamma_j(H_j,A_j;\psi_{j})+\sum_{k=j+1}^{K}\left[\gamma_k(H_k,\hat{d}_k^{\text{opt}};\hat{\psi}_{k})-\gamma_k(H_k,A_k;\hat{\psi}_{k})\right],
\end{aligned}
\end{equation*}
which is usually non-smooth in $\psi_j$ due to $d_j^{\text{opt}}=d_j^{\text{opt}}(H_j;\psi_j)=\arg\max_{a_j}\gamma_j(H_j,a_j;\psi_j)$. In summary, g-estimation of the optimal drSNMM, as specified by (\ref{opt_SNMM})-(\ref{EE_optSNMM}), allows us to learn the optimal treatment regime in a backward manner and the closed-form estimator is available in each stage when a linear blip function is assumed.

Although we described A-learning from two different perspectives (compare (\ref{Q-semi})-(\ref{response_A}) versus (\ref{opt_SNMM})-(\ref{EE_optSNMM})), they are essentially equivalent in terms of both modelling and estimation procedure. Note that under the sequential randomization assumption, the Q-contrast function in (\ref{Q-semi}) satisfies
\begin{equation*}
\begin{aligned}
C_j(h_j;\psi_j)=\mathbb{E}[Y^{\ast}(\bar{a}_{j-1},1,\underline{d}_{j+1}^{\text{opt}})-Y^{\ast}(\bar{a}_{j-1},0,\underline{d}_{j+1}^{\text{opt}})|H_j=h_j],
\end{aligned}
\end{equation*}
which, together with the definition of the optimal blip function in (\ref{opt_SNMM}), gives the relationship between $C_j(\cdot)$ and $\gamma_j(\cdot)$ as $
\gamma_j(H_j,A_j;\psi_j)=A_jC_j(H_j;\psi_j).
$
Therefore, modelling the contrast function is the same as modelling the blip function. As for estimation, EE (\ref{EE_K}) is a particular case of the general g-estimation (\ref{EE_optSNMM}) by giving a null model to $\mathbb{E}\left[U_{j}(\psi_j)|H_{j}\right]$ (i.e., setting $\mathbb{E}\left[U_{j}(\psi_j)\mid H_{j}\right]$ to zero) and setting $g_j(H_j,A_j)$ to be $A_j\partial C_{j}(H_{j};\psi_j)/\partial \psi_j$. Moreover, the updating of the response given by ($\ref{response_A}$) corresponds to the updating of $U_j(\psi_j)$ in (\ref{counter_outcome}).

In our simulated example, EE (\ref{EE_optSNMM}) for $\psi_2$  can be written as 
\begin{equation}\label{A_stage2_example}
\sum_{i=1}^{n}R_{2i}\left[Y_{i}-A_{2i}R_{2i}^{\top}\psi_2-D_{2i}^{\top}\xi_2\right]\left[A_{2i}-\mathbb{E}\left(A_{2i}\mid H_{2i};\hat{\alpha}_2\right)\right]=0,
\end{equation}
where $\hat{\alpha}_2$ is obtained by separately fitting a logistic regression model for $A_2$ on $(1,L_2)$, and $D_{2}^{\top}\xi_2$ specifies the term $\mathbb{E}\left[U_{2}(\psi_2)\mid H_{2}\right]=\mathbb{E}\left[Y-A_{2}R_{2}^{\top}\psi_2\mid H_{2}\right]$. Solving EE (\ref{A_stage2_example}) together with
\begin{equation*} 
\sum_{i=1}^{n}D_{2i}\left[Y_{i}-A_{2i}R_{2i}^{\top}\psi_2-D_{2i}^{\top}\xi_2\right]=0, 
\end{equation*}
leads to $\hat{\psi}_2=(677.83, -1.92)^{\top}$ and  $\hat{\xi}_2=(243.47, 1.51, 21.84, -0.48,  0.300)^{\top}$ (see the supplementary R markdown for the code). Then we have $\hat{d}_2^{\text{opt}}=I\left\{677.83-1.92L_2>0\right\}$, i.e., treat if $L_2<353$; and $\widehat{V}_{2}^{A}=Y+(I\{677.83-1.92L_2>0\}-A_2)(677.83-1.92L_2)$. As shown in Table \ref{tab:sim_data_A_stage2},  the optimal treatment actions in stage 2 for the three  individuals happen to be consistent with their treatments received in practice, and therefore their updated responses $\widehat{V}_{2}^{A}$'s are the same as $Y$'s. Estimation in stage 1 is performed by solving 
\begin{equation} \label{A_stage1_example}
\sum_{i=1}^{n}R_{1i}\left[\widehat{V}^{A}_{2i}-A_{1i}R_{1i}^{\top}\psi_j-D_{1i}^{\top}\xi_1\right]\left[A_{1i}-\mathbb{E}\left(A_{1i}\mid H_{1i};\hat{\alpha}_1\right)\right]=0, 
\end{equation}
\begin{equation*}
\sum_{i=1}^{n}D_{1i}\left[\widehat{V}^{A}_{2i}-A_{1i}R_{1i}^{\top}\psi_1-D_{1i}^{\top}\xi_1\right]=0, 
\end{equation*}
with $\hat{\alpha}_1$  obtained from fitting a logistic regression model for $A_1$ on $(1,L_1)$. We get $\hat{\psi}_1=(304.43,-1.13)^{\top}$ (and $\hat{\xi}_1=(307.81,1.78)^{\top}$), which leads to $\hat{d}_1^{\text{opt}}=I\left\{304.43-1.13L_1>0\right\}$, i.e., treat if $L_1<269$, and  $\widehat{V}_{1}^{A}=\widehat{V}_{2}^{A}+(I\{304.43-1.13L_1>0\}-A_1)(304.43-1.13L_1)$. As can be seen from this example, $\xi_j$, though estimated, is not involved in calculating $\widehat{V}_{j}^{A}$. The contribution of $\xi_j$ to $\hat{d}_j^{\text{opt}}$ and $\widehat{V}_{j}^{A}$ is indirect  through improving the efficiency of $\hat{\psi}_j$. See Section \ref{sec:simcase1} for further details. Moreover, the construction of $\widehat{V}_{j}^{A}$ given by (\ref{response_A}) determines the following monotonic relationship: $\widehat{V}_{1}^{A}\geq \widehat{V}_{2}^{A} \geq Y$ for all realizations of the history; this does not hold for Q-learning. The average values (over 500 individuals) of $\widehat{V}_{1}^{A}$ and $\widehat{V}_{2}^{A}$  are  $1122.99$ and $1067.92$, respectively.

\begin{table}[h]
\centering
\caption{\label{tab:sim_data_A_stage2}  The predicted optimal treatment actions and value functions from A-learning for the three random individuals in the simulated example.}
\begin{tabular*}{0.92\linewidth}{@{\extracolsep{\fill}} cccccccccc}
\hline \vspace{-0.3cm}\\
Individual & \textcolor{blue}{$L_1$} &\textcolor{blue}{$A_1$} & $L_2$ & $A_2$ & $Y$ & $\hat{d}_2^{\text{opt}}$ & \textcolor{blue}{$\widehat{V}_{2}^{A}$} &  $\hat{d}_1^{\text{opt}}$ & $\widehat{V}_{1}^{A}$ \\   \hline
1& \textcolor{blue}{356.03} & \textcolor{blue}{0} &357.76&  0& 1016.17 & 0 & \textcolor{blue}{1016.17} & 0 & 1016.17\\
2& \textcolor{blue}{477.55} &\textcolor{blue}{0} &546.19&  0& 1108.56  & 0 & \textcolor{blue}{1108.56} & 0 & 1108.56\\
3&\textcolor{blue}{324.65} &\textcolor{blue}{1} &330.79&  1&  898.41   & 1 & \textcolor{blue}{898.41} & 0 & 960.84\\
  \hline
\end{tabular*}
\end{table}

In addition to the contrast/blip-based A-learning, \cite{murphy2003optimal} proposed to estimate optimal treatment rules by directly modelling the regret functions, which is referred to as regret-based A-learning. Specifically, the regret function in the $j$th stage is defined as
\begin{equation}\label{regret_def}
\begin{aligned}
\mu_j(h_j,a_j)=\mathbb{E}[Y^{\ast}(\bar{a}_{j-1},d_{j}^{\text{opt}},\underline{d}_{j+1}^{\text{opt}})-Y^{\ast}(\bar{a}_{j-1},a_j,\underline{d}_{j+1}^{\text{opt}})|H_j=h_j], \quad j=1,\dots,K,
\end{aligned}
\end{equation}
which measures the expected loss in the final outcome by making decision $a_j$ rather than the optimal decision in stage $j$ among individuals with history $h_j$ when following the optimal treatment rules subsequently. Clearly, regret functions are conceptually equivalent to blip functions: $\mu_j(h_j,a_j)=\max_{a_j}\gamma_j(h_j,a_j)-\gamma_j(h_j,a_j)$ and  $\gamma_j(h_j,a_j)=\mu_j(h_j,0)-\mu_j(h_j,a_j)$. However, they are not equivalent in parameterization in the sense that a smooth blip function usually implies a non-smooth regret function due to the $\max$ operation \citep{moodie2007demystifying}. \cite{murphy2003optimal} proposed to model the regret functions by 
\begin{equation*}
\begin{aligned}
\mu_j(h_j,a_j)=\eta_j(h_j)f(a_j-d^{\text{opt}}_j(h_j)),
\end{aligned}
\end{equation*}
where $f(x)$ is a known `link' function which attains its minimal value 0 at $x=0$; and  $\eta_j(h_j)$ is a positive term indicating the price we need to pay for a sub-optimal decision rule at stage $j$. By specifying parsimonious parametric models for both $\eta_j(h_j)$ and $d^{\text{opt}}_j(h_j)$, we can obtain a parameterization of $\mu_j(h_j,a_j)$ whose parameters can be estimated via (i) the iterative minimization method developed by \cite{murphy2003optimal}, (ii) the regret-regression approach proposed by \cite{henderson2010regret} or (iii) g-estimation by deriving the corresponding contrast/blip functions.

\subsubsection{Causal tree}
The causal tree (CT) method \citep{athey2016recursive,blumlein2022learning} described in this section aims to extend the parametric specification $C_j(h_j;\psi_j)$ for the contrast function  in A-learning to a non-parametric, flexible but still interpretable tree-based model $C_j(h_j;\Pi_j)$ where $\Pi_j$ is a tree learned from the data for stage $j$. Specifically, $\Pi_j$ indicates a partition of the input space (here the space of $h_j$ in the $j$th stage) into cuboid regions (also referred to as leaves or terminal nodes) and is built recursively by identifying the splitting variable and its splitting value in each split. Let $C_j(h_j)$ denote the unknown true contrast function in the $j$th stage. Then for a given tree or partitioning $\Pi_j$, the tree-based contrast function $C_j(h_j;\Pi_j)$ is defined as
\begin{equation*}
\begin{aligned}
C_j(h_j;\Pi_j)&=\mathbb{E}[Y^{\ast}(\bar{a}_{j-1},1,\underline{d}_{j+1}^{\text{opt}})-Y^{\ast}(\bar{a}_{j-1},0,\underline{d}_{j+1}^{\text{opt}})\mid H_j\in \ell(h_j,\Pi_j)]\\
&=\mathbb{E}[C_j(H_j)\mid H_j\in \ell(h_j,\Pi_j)],
\end{aligned}
\end{equation*}
where $\ell(h_j,\Pi_j)$ denotes the leaf $\ell \in \Pi_j$ such that $h_j \in \ell$. That is, $C_j(h_j;\Pi_j)$ is the average of  $C_j(h_j)$ over the leaf to which $h_j$ belongs. Therefore, $C_j(h_j;\Pi_j)$ 
can be viewed as a multi-dimensional step-function (leaf-wise) approximation to $C_j(h_j)$ \citep{athey2016recursive}, i.e., $C_j(h_j;\Pi_j)=C_j(h^\prime_j;\Pi_j)$ if $\ell(h_j,\Pi_j)=\ell(h^\prime_j,\Pi_j)$; and CT is in fact a non-parametric A-learning approach. Compared with $C_j(h_j;\psi_j)$ where a parametric model is prespecified, CT allows the model, here the tree $\Pi_j$, to be learned from the data, and is thus more flexible, especially in dealing with multivariate $L_j$.  As for the construction of $\Pi_j$, \cite{athey2016recursive} developed an honest splitting criterion which rewards a split leading to improved heterogeneity in the contrast function and at the same time penalizes the increased variances in leaf estimates. We omit the details on how to build a causal tree. Please refer to \cite{athey2016recursive} for the splitting criterion in detail and the difference between the conventional regression tree and the causal tree.

Suppose that $\Pi_j$ in stage $j$ (also $j+1,\dots,K$) has already been learned from the data, then $C_j(h_j;\Pi_j)$ can be estimated from a sample $\mathcal{S}$ by
\begin{equation}
\begin{aligned}\label{contrast_estimate_CT}
\widehat{C}_j(h_j;\mathcal{S},\Pi_j)=\frac{1}{\# \mathcal{S}_1(h_j,\Pi_j)}\sum_{i \in \mathcal{S}_1(h_j,\Pi_j)}\widehat{V}^{CT}_{j+1,i}-\frac{1}{\# \mathcal{S}_0(h_j,\Pi_j)}\sum_{i \in \mathcal{S}_0(h_j,\Pi_j)}\widehat{V}^{CT}_{j+1,i},
\end{aligned}
\end{equation}
where ${S}_a(h_j,\Pi_j)=\{i \in \mathcal{S}: H_{ji}\in \ell(h_j,\Pi_j) \land  A_{ji}=a\}$ for $a=0, 1$, and 
\begin{equation*}
\begin{aligned}
\widehat{V}^{CT}_{j+1}=Y+\sum_{k=j+1}^{K}\left(I\left\{\widehat{C}_k(H_k; \mathcal{S},\Pi_k)>0\right\}-A_k\right)\widehat{C}_k(H_k;\mathcal{S},\Pi_k),
\end{aligned}
\end{equation*}
for $j=1,\dots,K$ with $\widehat{V}^{CT}_{K+1}$ defined as $Y$. Note that (\ref{contrast_estimate_CT}) simply compares the treated and untreated units within the leaf $ \ell(h_j,\Pi_j)$, which in fact assumes the leaf $ \ell(h_j,\Pi_j)$ is small enough that the responses $\widehat{V}^{CT}_{j+1,i}$ within the leaf are roughly identically distributed \citep{wager2018estimation}. Otherwise, modification based on propensity scores is required, for example, using propensity score stratification or inverse probability weighting to correct for variations in propensity within the leaf \citep{powers2018some,athey2016recursive}. 

\begin{figure}
\centering
\includegraphics[height=6cm,width=14cm]{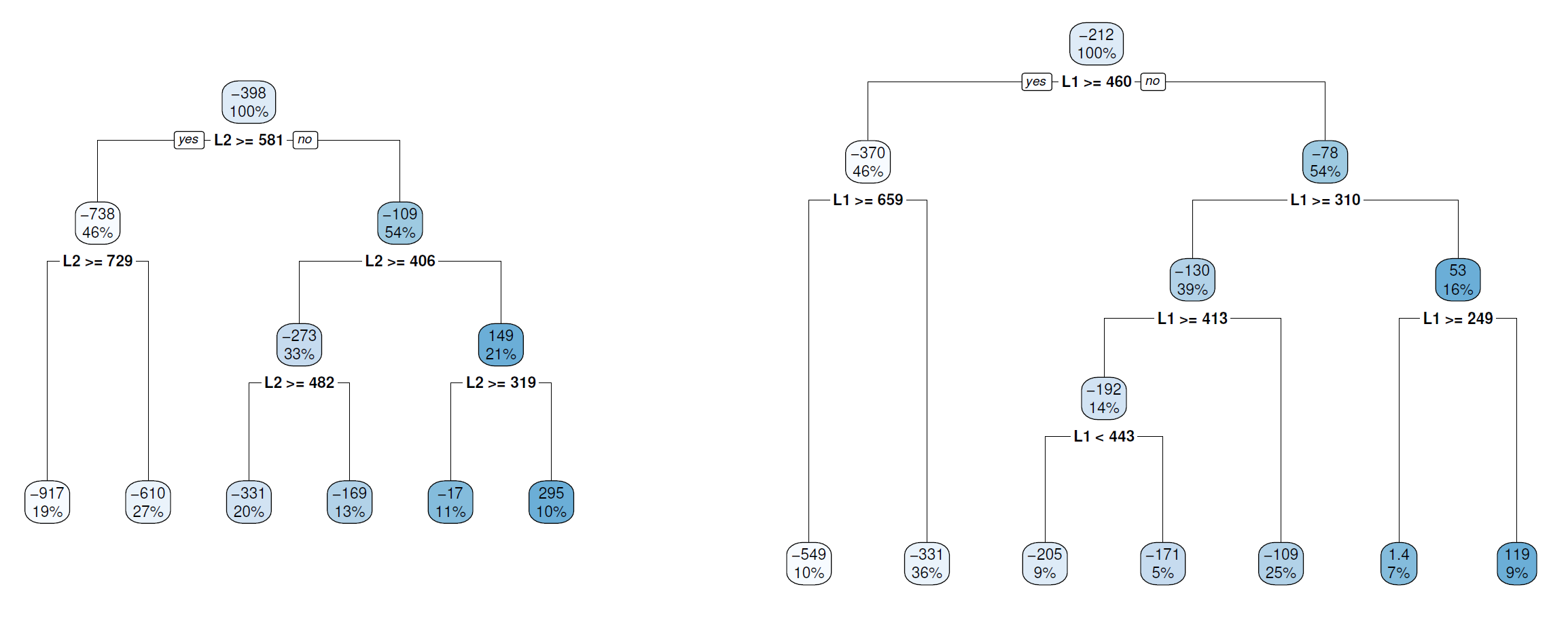}
\caption{\label{fig:sim_data_tree}  The estimated contrast function $\widehat{C}_j(\cdot;\Pi_j)$ indicated by the values in the terminal nodes and the 
corresponding tree  $\Pi_j$ learned from the data for $j=2$ (left) and $j=1$ (right).}
\end{figure}

Returning to the simulated example, we learn the tree structure $\Pi_2$ with the honest splitting rule \citep{athey2016recursive} and estimate the tree-based contrast function $C_2(\cdot;\Pi_2)$ with the inverse probability of treatment weights (IPTW) \citep{wallace2015doubly} for modification.  The estimated contrast function $\widehat{C}_2(\cdot; \Pi_2)$ is given by the values inside the terminal nodes in the left subfigure of Figure \ref{fig:sim_data_tree} and the tree structure $\Pi_2$ is indicated by the splitting variable (here $L_2$ alone) and the splitting values, making up a partition of the feature space in terms of the magnitude of the contrast function in stage 2. Given $\widehat{C}_2(\cdot; \Pi_2)$, it is straightforward to derive $\hat{d}_2^{\text{opt}}$ and $\widehat{V}^{CT}_{2}$ for each individual. See Table \ref{tab:sim_data_tree_stage2} for the predicted $\hat{d}_2^{\text{opt}}$'s and  $\widehat{V}^{CT}_{2}$'s for the three individuals. Estimation in stage 1 is performed similarly except that  $\widehat{V}^{CT}_{2}$ serves as the response variable and $L_1$ constitutes the feature space. The estimated contrast function is illustrated in the right subfigure of Figure \ref{fig:sim_data_tree} and the predicted  $\hat{d}_1^{\text{opt}}$'s and  $\widehat{V}^{CT}_{1}$'s are included in Table \ref{tab:sim_data_tree_stage2}. Moreover, the average values (over 500 individuals) of $\widehat{V}_{1}^{CT}$ and $\widehat{V}_{2}^{CT}$  are  $1129.93$ and $1069.95$, respectively.

\begin{table}
\centering
\caption{\label{tab:sim_data_tree_stage2}  The predicted optimal treatment actions and value functions from causal tree method for the three random individuals in the simulated example.}
\begin{tabular*}{0.95\linewidth}{@{\extracolsep{\fill}} cccccccccc}
\hline \vspace{-0.3cm}\\
Individual & \textcolor{blue}{$L_1$} &\textcolor{blue}{$A_1$} & $L_2$ & $A_2$ & $Y$ & $\hat{d}_2^{\text{opt}}$ & \textcolor{blue}{$\widehat{V}_{2}^{CT}$} &  $\hat{d}_1^{\text{opt}}$ & $\widehat{V}_{1}^{CT}$ \\   \hline
1& \textcolor{blue}{356.03} & \textcolor{blue}{0} &357.76&  0& 1016.17 & 0 & \textcolor{blue}{1016.17} & 0 & 1016.17\\
2& \textcolor{blue}{477.55} &\textcolor{blue}{0} &546.19&  0& 1108.56  & 0 & \textcolor{blue}{1108.56} & 0 & 1108.56\\
3&\textcolor{blue}{324.65} &\textcolor{blue}{1} &330.79&  1&  898.41   & 0 & \textcolor{blue}{915.11} & 0 & 1023.82\\
  \hline
\end{tabular*}
\end{table}

As a summary of the aforementioned indirect methods, Algorithm \ref{algorithm}  outlines the estimation procedures of Q-learning, A-learning and causal tree. All of them can be implemented through a unified backward induction.

\begin{algorithm}
	\caption{\label{algorithm} Estimate optimal dynamic treatment regime via backward induction} 
	\begin{algorithmic}
            \State $\widehat{V}_{K+1}:=Y$
		\For {$j=K,\ldots,1$}
             \State Estimate the Q-function or Q-contrast function in the $j$th stage:
                \begin{itemize}
                    \item if Q-learning: estimate the full Q-function or equivalently $\theta_j=(\psi_j,\xi_j)$ by fitting (\ref{Q-full}) for $\widehat{V}_{j+1}$ via OLS;
                    \item if A-learning: 
                    \begin{itemize} 
                    \item parametric: estimate the parametric Q-contrast function or equivalently $\psi_j$ by solving (\ref{EE_optSNMM});
                    \item non-parametric (causal tree): estimate the causal tree structure by the honest estimation algorithm and derive the estimated tree-based contrast via (\ref{contrast_estimate_CT}).
                    \end{itemize}
                \end{itemize}
           \State Derive the optimal treatment rule: $\hat{d}_{j}^{\text{opt}}(H_j)=I(\hat{C}_j(H_j)>0)$. 
           \State Update the response variable: 
           \begin{itemize}
           \item if Q-learning, $\widehat{V}_{j}^{Q}=I\left\{C_j(H_j;\hat{\psi}_j)>0\right\}C_j(H_j;\hat{\psi}_j)+m_j(H_j;\hat{\xi}_j)$
            \item if A-leaning,  $\widehat{V}_{j}^{A}=\widehat{V}^{A}_{j+1}+\left(\hat{d}_{j}^{\text{opt}}(H_j)-A_j\right)\hat{C}_j(H_j)$
           
           \end{itemize}

		\EndFor
            
	\end{algorithmic} 
        \Return $\hat{d}^{\text{opt}}=(\hat{d}^{\text{opt}}_1,\dots,\hat{d}^{\text{opt}}_K)$
\end{algorithm}

\begin{figure}
\centering
\includegraphics[height=3.3cm,width=14cm]{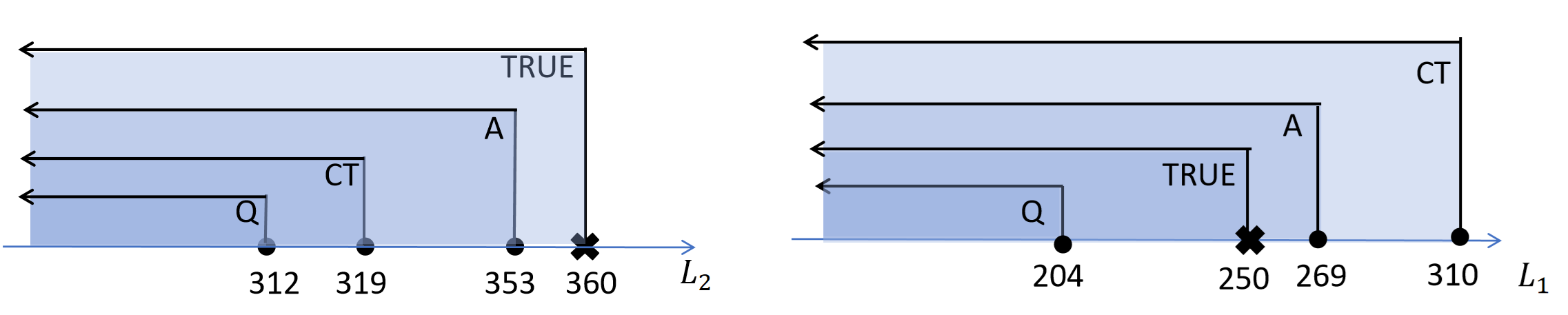}
\caption{\label{fig:example} A summary of the estimated decision rules (to whom the treatment should be assigned in stage 2 (left) and stage 1 (right)) given by Q-learning, A-learning and causal tree, in comparison with the true rules.}
\end{figure}

\subsection{Direct methods}
A direct method focuses on the parameterization of the class of treatment regimes $\mathcal{D}$ and solves the regime-related optimization problem defined by (\ref{1}). Under the identification conditions given in Section \ref{sec:assump}, the  counterfactual marginal mean $\mathbb{E} [Y^{\ast}(d)]$  can be expressed in terms of the observed data by applying (augmented) inverse probability weighting. This results in a discontinuous objective function due to the involvement of the indicator function. \cite{zhang2012robust,zhang2013robust} directly worked on the resulting objective function and developed the inverse probability weighted estimator (IPWE) and the augmented inverse probability weighted estimator (AIPWE), depending on how $\mathbb{E} [Y^{\ast}(d)]$ was estimated. \cite{zhao2012estimating,zhao2015new} transformed the maximization problem into a weighted classification problem and substituted the 0-1 loss with a convex surrogate loss, specifically the hinge loss used notably for support vector machines (SVMs), for computational tractability.  This weighted classification framework is termed outcome weighted learning (OWL), and includes several variants, such as the backward OWL (BOWL) and simultaneous OWL (SOWL) based on different learning approaches, as well as the augmented outcome-weighted learning (AOL) in \cite{liu2018augmented} that adopted a robust augmentation to the original weights in BOWL, and the recently proposed smooth Fisher consistent surrogate loss optimization \citep{laha2024finding} which replaced the hinge loss in SOWL by a Fisher consistent surrogate loss under the DTR setting. We will take (A)IPWE and BOWL as representative methods of the two general frameworks (i.e. working directly with $\mathbb{E} [Y^{\ast}(d)]$ and recasting as a weighted classification problem) to illustrate direct methods in this section.

\subsubsection{Augmented inverse probability weighted estimator}
Instead of deriving the optimal dynamic treatment regimes from the estimated Q-functions or contrast functions, \cite{zhang2012robust,zhang2013robust} proposed to work on a 
specified class of regimes $\mathcal{D}_{\psi}$ with elements $d_{\psi}=(d_{\psi_{1}},\dots,d_{\psi_{K}})$ indexed by $\psi$, and obtain the optimal one $d^{\text{opt}}=d_{\psi^{\text{opt}}}$ by estimating $ \psi^{\text{opt}}=\arg\max_{\psi}\mathbb{E}[Y^{\ast}(d_\psi)]$. Depending on the approach used to estimate $\mathbb{E}[Y^{\ast}(d_\psi)]$, they developed inverse probability weighted estimator (IPWE) and augmented inverse probability weighted estimator (AIPWE). 

Let $J_{\psi}$ be a discrete variable taking values $1,\dots,K,\infty$ corresponding to the extent to which the observed sequential treatments are consistent with the specified regime $d_\psi$. Specifically, $J_\psi=1$ if $A_1\neq d_{\psi_1}(L_1)$; $J_\psi=2$ if  $A_1= d_{\psi_1}(L_1)$ but $A_2\neq d_{\psi_2}(H_2)$ (here $d_{\psi_2}(H_2)$ can also be written as $d_{\psi_2}(\bar{L}_2)$ since $A_1= d_{\psi_1}(L_1)$) ; $J_\psi=K$ if  $A_j= d_{\psi_j}(H_j)$ for $j=1,\dots,K-1$ but  $A_K\neq d_{\psi_K}(H_K)$; and $J_\psi=\infty$ if $A_j= d_{\psi_j}(H_j)$ for $j=1,\dots,K$. Accordingly, define the $\lambda_{\psi,j}(\bar{L}_{j})$ as the probability of failing to match $d_\psi$ at stage $j$ conditional on being consistent with $d_\psi$ through all prior $j-1$ stages, i.e., $\lambda_{\psi,j}(H_{j})=P(A_j\neq d_{\psi_j}(H_j)|\bar{L}_j,\bar{A}_{j-1}=\bar{d}_{\psi_{j-1}}(H_{j-1}))$ and  $\lambda_{\psi,j}(H_{j})$ can be rewritten as $\lambda_{\psi,j}(\bar{L}_{j})$  due to similar reasons mentioned above. Let $\pi_{j}(\bar{l}_{j},\bar{a}_{j-1})=P(A_j=1|\bar{L}_j=\bar{l}_j,\bar{A}_{j-1}=\bar{A}_{j-1})$ be the probability of receiving the treatment in stage $j$ given the history. Then we have 
\begin{equation*}
\begin{aligned}
\lambda_{\psi,j}(\bar{L}_{j})=\pi_{j}\left(\bar{L}_{j},\bar{d}_{\psi_{j-1}}(\bar{L}_{j-1})\right)^{1-d_{\psi_j}(\bar{L}_j)}\times \left[1-\pi_{j}\left(\bar{L}_{j},\bar{d}_{\psi_{j-1}}(\bar{L}_{j-1})\right)\right]^{d_{\psi_j}(\bar{L}_j)},
\end{aligned}
\end{equation*}
and the probability of being consistent with $d_\psi$ through at least the $j$th stage can be expressed in terms of $\lambda_{\psi,j}(\bar{L}_{j})$ as the survival probability in a discrete-time survival model:
\begin{equation*}
\begin{aligned}
M_{\psi,j}(\bar{L}_j):=P(J_{\psi}>j|\bar{L}_{j})=\prod_{k=1}^{j}(1-\lambda_{\psi,k}(\bar{L}_{k})).
\end{aligned}
\end{equation*}

Then the IPWE and AIPWE can be formalized with the preceding notations. Note that
\begin{equation*}
\begin{aligned}
\mathbb{E}[Y^{\ast}(d_\psi)]=\mathbb{E}\left[\frac{ \prod_{j=1}^{K}I\left\{A_j=d_{\psi_j}(H_j)\right\} }
{ \prod_{j=1}^{K}P(A_j|H_j) }Y\right]
\end{aligned}
\end{equation*}
under the assumptions in Section \ref{sec:assump}. It is thus straightforward to estimate $\mathbb{E}[Y^{\ast}(d_\psi)]$ for any fixed $\psi$ by 
\begin{equation*}
\begin{aligned}
\frac{1}{n}\sum_{i=1}^{n}\frac{I(J_{\psi,i}=\infty)}{M_{\psi,K}(\bar{L}_{Ki})}Y_i,
\end{aligned}
\end{equation*}
where only individuals whose observed sequential treatments are consistent with the $\psi$-indexed regime $d_{\psi}$ are included, weighted by the inverse of the probability of being consistent through all stages. Given a consistent estimate of $M_{\psi,K}(\bar{L}_{K})$ (or equivalently $\lambda_{\psi,j}(\bar{L}_{j})$ for $j=1,\dots,K$), e.g., $\hat{M}_{\psi,K}(\bar{L}_{K})$, IPWE is defined by
\begin{equation}\label{IPWE}
\begin{aligned}
\text{IPWE}(\psi)=\frac{1}{n}\sum_{i=1}^{n}\frac{I(J_{\psi,i}=\infty)}{\hat{M}_{\psi,K}(\bar{L}_{Ki})}Y_i,
\end{aligned}
\end{equation}
and the optimal regime in $\mathcal{D}_{\psi}$ is obtained by finding $\psi^{\text{opt}}$ which maximizes (\ref{IPWE}) over $\psi$. Augmenting the IPWE by a mean-zero term yields the following AIPWE
\begin{equation}\label{AIPWE}
\begin{aligned}
\text{AIPWE}(\psi)=\frac{1}{n}\sum_{i=1}^{n}\left\{\frac{I(J_{\psi,i}=\infty)}{\hat{M}_{\psi,K}(\bar{L}_{Ki})}Y_i+\sum_{j=1}^{K}\frac{I(J_{\psi,i}=j)-\hat{\lambda}_{\psi,j}(\bar{L}_{ji})I(J_{\psi,i}\geq j)}{\hat{M}_{\psi,j}(\bar{L}_{ji})}\hat{\omega}_{\psi,j}(\bar{L}_{ji})\right\},
\end{aligned}
\end{equation}
where $\hat{\omega}_{\psi,j}(\bar{l}_{j})$ is an estimate of $\omega_{\psi,j}(\bar{l}_{j})=\mathbb{E}[Y^{\ast}(d_{\psi})|\bar{L}_{j}^{\ast}(\bar{d}_{\psi_{j-1}})=\bar{l}_{j}]$. Compared with IPWE, AIPWE gains efficiency through the augmentation term in (\ref{AIPWE}) and is doubly robust in the sense that (\ref{AIPWE}) is a consistent estimator of $\mathbb{E}[Y^{\ast}(d_\psi)]$ if either the $\hat{\lambda}_{\psi,j}(\bar{L}_{j})$ or $\hat{\omega}_{\psi,j}(\bar{l}_{j})$, for $j=1,\dots,K$, is correctly specified. However, it is challenging to exactly model and estimate $\omega_{\psi,j}(\bar{l}_{j})$ since it is itself $\psi$-dependent, which means the model of $\omega_{\psi,j}(\bar{l}_{j})$ has to be fitted at each value of $\psi$ encountered in the optimization of $\text{AIPWE}(\psi)$. A pragmatic method, proposed in \cite{zhang2013robust}, is to replace $\hat{\omega}_{\psi,j}(\bar{L}_{j})$ in (\ref{AIPWE}) by $Q_j(\bar{L}_j,\bar{d}_{\psi,j}(\bar{L}_j);\hat{\theta}_j)$, i.e., the estimated Q-functions obtained via Q-learning. While $Q_j(\bar{l}_j,\bar{d}_{\psi,j}(\bar{l}_j);\hat{\theta}_j)$ is an estimate of $\mathbb{E}[Y^{\ast}(\bar{d}_{\psi_j},\underline{d}^{\text{opt}}_{j+1})|\bar{L}_{j}^{\ast}(\bar{d}_{\psi_{j-1}})=\bar{l}_{j}]$ rather than $\mathbb{E}[Y^{\ast}(d_{\psi})|\bar{L}_{j}^{\ast}(\bar{d}_{\psi_{j-1}})=\bar{l}_{j}]$, it still improves efficiency compared with IPWE and will be close to  $\omega_{\psi,j}(\bar{l}_{j})$ when $\psi$ takes value around $\psi^{\text{opt}}$.  Moreover, the computational burden has been greatly alleviated since the estimation of $\hat{\theta}_j$, as described in Section \ref{sec:Q-learning}, is independent of $\psi$. Applying IPWE and AIPWE to our simulated example, we got the estimated optimal DTR $(I\{L_1<354\},I\{L_2<424\})$ from IPWE and $(I\{L_1<268\},I\{L_2<356\})$ from AIPWE. See the supplementary R markdown for the computational details.

\subsubsection{Backward outcome weighted learning}
Having indicator functions in
the objective functions, (\ref{IPWE}) and (\ref{AIPWE}), results in significant challenges to the optimization of these objective functions over $\psi$ due to their discontinuous and non-differentiable nature.  OWL overcomes these optimization challenges by replacing the 0-1 loss with a surrogate loss after recasting the original maximization problem to an equivalent classification  problem. We illustrate the idea of OWL by taking BOWL as an example.

Consider stage $j$ and assume $d^{\text{opt}}_{j+1},\dots, d^{\text{opt}}_{K}$ are known. Then searching for $d^{\text{opt}}_{j}$ requires maximizing
\begin{equation}\label{owl_max}
\begin{aligned}
\mathbb{E}\left[\frac{Y \prod_{k=j+1}^{K}I\left\{A_k=d^{\text{opt}}_{k}(H_k)\right\}}
{ \prod_{k=j}^{K}P(A_k\mid H_k) } I\left\{A_j=d_{\psi_j}(H_j)\right\}  \middle| H_j\right]
\end{aligned}
\end{equation}
over $\psi_j$, which is equivalent to minimizing
\begin{equation}\label{owl_min}
\begin{aligned}
\mathbb{E}\left[\frac{Y \prod_{k=j+1}^{K}I\left\{A_k=d^{\text{opt}}_{k}(H_k)\right\}}
{ \prod_{k=j}^{K}P(A_k\mid H_k) } I\left\{A_j\neq d_{\psi_j}(H_j)\right\}  \middle| H_j\right],
\end{aligned}
\end{equation}
since the sum of (\ref{owl_max}) and (\ref{owl_min}):
\begin{equation*}
\begin{aligned}
\mathbb{E}\left[\frac{Y \prod_{k=j+1}^{K}I\left\{A_k=d^{\text{opt}}_{k}(H_k)\right\}}
{ \prod_{k=j+1}^{K}P(A_k\mid H_k) }\frac{1}{P(A_j\mid H_j)} \middle| H_j\right],
\end{aligned}
\end{equation*}
is free of $\psi_j$. By observing that (\ref{owl_min}) can be viewed as a  classification problem under 0-1 loss weighted by $Y \prod_{k=j+1}^{K}I\left\{A_k=d^{\text{opt}}_{k}(H_k)\right\}/
{ \prod_{k=j}^{K}P(A_k\mid H_k) }$,  \cite{zhao2012estimating,zhao2015new} tackle the optimization difficulty in (\ref{owl_min}), resulting from the indicator function $I\left\{A_j\neq d_{\psi_j}(H_j)\right\}$, by replacing this 0-1 loss with the hinge loss employed in SVM. Specifically, let $d_{\psi_j}(H_j)=\text{sign}(\phi_j(H_j;\psi_j))$, where $\phi_j(\cdot;\psi_j)$ is the decision function at the $j$th decision point and the treatment rule is determined by the sign of the decision function. Following the notation convention in the classification setting, define $\tilde{A}_j=2A_j-1\in\{-1,1\}$ for $j=1,\dots,K$. Then the (empirical) objective function becomes
\begin{equation}\label{hinge_loss}
\begin{aligned}
\frac{1}{n}\sum_{i=1}^{n}\frac{Y_i \prod_{k=j+1}^{K}I\left\{\tilde{A}_{ki}=d^{\text{opt}}_{k}(H_{ki})\right\}}
{ \prod_{k=j}^{K}\hat{P}(\tilde{A}_{ki}\mid H_{ki}) } [1-\tilde{A}_{ji}\phi_j(H_{ji};\psi_j)]_{+}+c_{j,n}\|\psi_j\|_2^{2},
\end{aligned}
\end{equation}
where $[x]_{+}=\max\{x,0\}$ and $c_{j,n}$ is the tuning parameter to balance the classification error and the complexity of the decision function. The optimization of (\ref{hinge_loss}) over $\psi_j$ then can proceed similarly to what is done with SVM. Moreover, using the developed  kernel trick in SVM,  the decision function $\phi_j$ can be defined and explored in a broader space, i.e., a reproducing kernel Hilbert space, rather than a restricted (with prespecified basis functions) parametric family index by $\psi_j$. This makes OWL suitable to handle nonlinear decision functions. Additionally, the hinge loss used in (\ref{hinge_loss}) suggests that the estimated decision function is determined by the support points in SVM, making OWL more robust to model misspecification but at the same time less efficient. We illustrate the implementation of OWL in Section \ref{sec:simcase2} where a simulated setting with multiple biomarkers are considered.

\begin{figure}
\centering
\includegraphics[height=7cm,width=14cm]{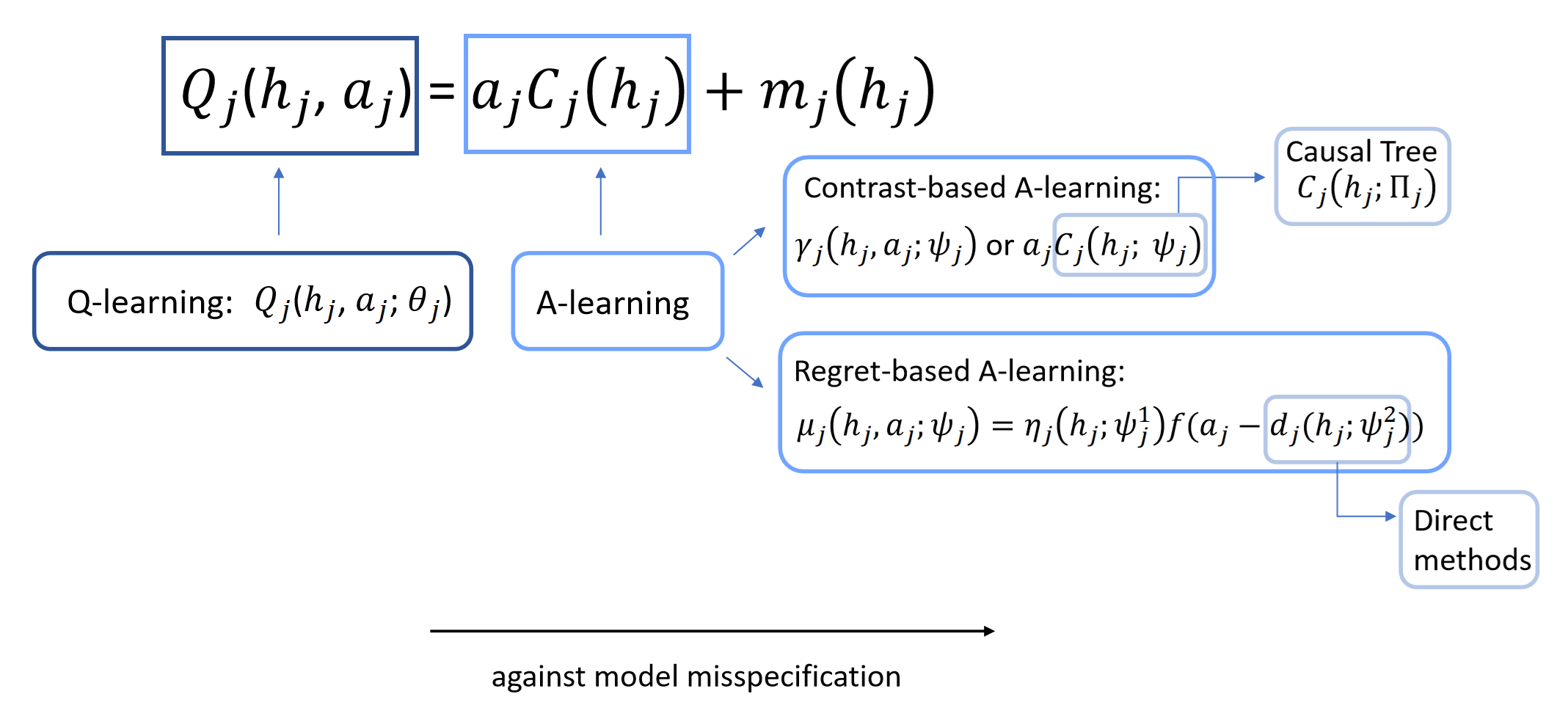}
\caption{\label{fig:diagram}  A relationship diagram of different methods in terms of model specification.}
\end{figure}

We provide Figure \ref{fig:diagram} to illustrate the model relationship between the aforementioned methods. Briefly speaking, indirect methods step towards robustness by positing more flexible models on some aspect of the conditional distribution (here, conditional mean function). Direct methods, e.g., (A)IPWE and OWL, achieve robustness by directly focusing on the specification of candidate treatment regimes. Furthermore, we present Table \ref{tab:model_summary} as a summary of these methods in terms of both modelling and estimation. The central term for model specification in Table \ref{tab:model_summary} refers to the term which must be correctly specified in order to get a consistent estimate of $d^{\text{opt}}$ under a given method, whereas the auxiliary term (also referred to as the nuisance working model) is the term which is not directly of
interest but must be considered in the estimation. All indirect methods, i.e., Q-learning, A-learning, and causal tree, allow stage-wise estimation, which facilitates checking the goodness of fit of the specified model. Closed-form solutions of $\psi_j$ can be found when the Q-function and the contrast function are linear in  $\psi_j$ in Q-learning and A-learning, respectively, and $\psi_j$'s are not shared between stages. Direct methods specify models on regimes themselves. For example, $d_{\psi_j}$ in (A)IPWE can be modelled as $I\{L_j>\psi_j\}$ if there is sufficient prior knowledge suggesting the treatment should be assigned when $L_j$ gets higher; whereas the modelling of $d_{\psi_j}$ in OWL follows the convention in classification setting, i.e., through modelling the decision function (boundary).  

Finally, it is worth noting that classifying DTR methods into indirect and direct approaches is rather general and coarse. Many recently developed methods are hybrid, seeking to utilize the strength of both approaches. For example, C-learning \citep{zhang2018c} works on a weighted classification problem based on the estimated contrast functions where the magnitude of the contrast function serves as the weight and the sign of the contrast function creates the label. \cite{zhang2018interpretable} proposed to firstly fit non-parametric Q-functions and then work on a maximization problem based on the estimated Q-functions over a class of list-based decision rules which are immediately interpretable to domain experts. In this way, they decoupled the interpretability of the estimated decision rules from that of the Q-functions. See also \cite{laber2015tree} and \cite{tao2018tree} for estimating decision rules which are representable as decision trees. Moreover, these rule-based methods \citep{zhang2018interpretable,laber2015tree,tao2018tree} are also flexible in dealing with categorical or continuous treatments at each stage.

\begin{table}
\centering
\caption{\label{tab:model_summary} A summary of the four methods introduced in Section \ref{sec:model}.} 
\begin{tabular}{c c c c c}
  \hline \vspace{-4mm} \\ 
  & \multicolumn{2}{c}{model specification}  & \multicolumn{2}{c}{estimation}\\
\cmidrule(rl){2-3} \cmidrule(rl){4-5}
\multicolumn{1}{c}{method}& \multicolumn{1}{c}{central} & \multicolumn{1}{c}{auxiliary}  & stage-wise & closed-form\\ \hline \vspace{-2mm} \\ 
Q & Q-functions &  null & $\checkmark^{1}$ & $\checkmark^{2}$ \vspace{2mm}\\ 
{A} & Q-contrast functions &  \makecell{\small{propensity or/$\text{and}$}\\ \small{treatment-free term}}  & $\checkmark^{1}$ &$\checkmark^{2}$ \vspace{2mm}\\ 
CT & $\text{null}^{3}$  & propensity & $\checkmark$ &  $\times$ \vspace{2mm} \\
\small{(A)IPWE} & \makecell{\small{parameterization of} \\ \small{candidate regimes  $d_{\psi}$}} &  \makecell{\small{propensity (and }\\ \small{$\text{Q-functions}^{4}$)}}  & $\times$ & $\times$ \\
BOWL &decision functions $\phi_j$'s&  propensity   & $\checkmark$ & $\times$ 
\\ \hline
\end{tabular}
\begin{tablenotes}\footnotesize
\item $^1$ if parameters are not shared between stages in Q-functions (Q-contrast functions) for Q-learning (A-learning).
\item $^2$ if Q-function (contrast function) in each stage is linear in parameters for Q-learning (A-learning).
\item $^3$ hyperparameters required in order to control the tree complexity.
\item $^4$ Q-functions work as approximations to $\mathbb{E}[Y^{\ast}(d_{\psi})|\bar{L}_{j}^{\ast}(\bar{d}_{\psi_{j-1}})=\bar{l}_{j}]$ for $j=1,\dots,K$ in the augmentation term.
\end{tablenotes}
\end{table}

\section{Simulation} \label{sec:sim}
To provide general advice to readers on choosing an appropriate method to estimate optimal DTRs, we compare the different models and estimation methods through simulation studies. However in causal inference with sequential treatments, specifying a data-generating model whose derived causal model is exactly the one we want is not straightforward, especially when the causal model of interest is a marginal structural model \citep{evans2023parameterizing,seaman2023simulating,robins1997estimation}.  When our focus is restricted to SNMMs with a continuous outcome $Y$, \cite{murphy2003optimal} showed that the conditional mean of $Y$ given $(\bar{L}_{K},\bar{A}_K)$ can be expressed in terms of the regret functions (or equivalently blip functions) as follows:
\begin{equation}
    E[Y|\bar{L}_{K},\bar{A}_K]=\mu_0+\sum_{j=1}^{K}\phi_{j}(\bar{L}_{j},\bar{A}_{j-1})-\sum_{j=1}^{K}\mu_{j}(H_j,A_j),\label{data_gene}
\end{equation}
where $\mu_0=\mathbb{E}[V_{1}(L_1)]=\mathbb{E}[Y^{\ast}(d^{\text{opt})})]$ is the marginal mean of the potential outcome under $d^{\text{opt}}$; $\phi_{j}(\bar{L}_{j},\bar{A}_{j-1})=V_j(\bar{L}_{j},\bar{A}_{j-1})-Q_{j-1}(\bar{L}_{j-1},\bar{A}_{j-1})$ for $j=1,...,K$, are mean-zero terms as indicated by (\ref{5}); and $\mu_{j}(H_j,A_j)$ for $j=1,...,K$ are regret functions as defined in (\ref{regret_def}). Moreover, the nuisance components in the joint distribution of $(L_1,A_1,\dots, L_K,A_K,Y)$, i.e., $\mu_0$, $\phi_{j}(\bar{L}_{j},\bar{A}_{j-1})$, the distribution of $L_j|\bar{L}_{j-1},\bar{A}_{j-1}$ and that of $A_j|\bar{L}_{j},\bar{A}_{j-1}$, and other aspects of the distribution $Y|\bar{L}_{K},\bar{A}_K$ except for the conditional mean, can be specified flexibly and without implicit constraints from the regret functions. We therefore adopt (\ref{data_gene}) in generating data and perform simulation studies to compare different models and estimation methods.  The R code for the simulations is provided in the supplementary R markdown document.

\subsection{Two decision points with one-dimensional $L_j$}\label{sec:simcase1}
In this section, the observed data are generated as follows:
$L_1\sim N(450,150^2)$,
$A_1|L_1\sim \text{Bin}(1,p_1)~\text{with}~ p_1=\text{expit}(2-0.06L_1)$; $L_2|L_1,A_1\sim N(1.25L_1,60^2)$,
$A_2|\bar{L}_2,A_1\sim \text{Bin}(1,p_2)~\text{with}~p_2=\text{expit}(0.8-0.04L_2);$ and 
$$Y|\bar{L}_2,\bar{A}_2 \sim N(400+1.6L_1,60^2)-\mu_1(H_1,A_1;\psi_1^{\ast})-\mu_2(H_2,A_2;\psi_2^{\ast}),$$
where $\mu_j(H_j,A_j;\psi_j^{\ast})=\left(I\{\psi_{j0}^{\ast}+\psi_{j1}^{\ast}L_j>0\}-A_j\right)(\psi_{j0}^{\ast}+\psi_{j1}^{\ast}L_j)$ for $j=1,2$ with values of $\psi_j^{\ast}$ given in Table \ref{tab:sim1_res}.  The true blip functions in this case are $\gamma_{j}(H_j,A_j;\psi_j^{\ast})=A_j(\psi_{j0}^{\ast}+\psi_{j1}^{\ast}L_j)$ for $j=1,2$ and $\mathbb{E}[Y^{\ast}({d^\text{opt}})]=400+1.6\times 450=1120$. The above setting is designed to mimic a longitudinal antiretroviral therapy study for HIV-infected patients with $L_j$ and $Y$ being the intermediate and final CD4 counts, respectively. Similar settings have been investigated in \cite{moodie2007demystifying}, \cite{zhang2013robust} and \cite{schulte2014q}. We performed Q-learning, A-learning and (A)IPWE in estimating $\psi^{\ast}=(\psi_{1}^{\ast},\psi_{2}^{\ast})$ and the estimation results are given in Table \ref{tab:sim1_res}. 

For Q-learning, the  Q-functions in stage 2 and 1 are specified as (\ref{Q2_example}) and (\ref{Q1_example}), respectively.
Since the true Q-functions, as indicated by the data-generating mechanism, involve nonlinear treatment-free terms, the posited Q-functions in (\ref{Q2_example}) and (\ref{Q1_example}) are therefore misspecified in both stages. Working with misspecified Q-functions leads to biased estimates of $\psi^{\ast}$ as shown in Table \ref{tab:sim1_res}. 

For A-learning, we restrict our attention to the case that both the blip functions (equivalently, the contrast functions) and the parametric working model for $\mathbb{E}[A_j\mid H_j]$ (i.e., the model for $\mathbb{E}\left[g_j(H_{j},A_{j})\mid H_{j}\right]$ in (\ref{EE_optSNMM})), say, $\mathbb{E}\left(A_{j}\mid H_{j};\alpha_j\right)$ are correctly specified, which ensures that the consistent estimate of $\psi_j^{\ast}$ is obtained by solving equation (\ref{EE_optSNMM}). For the choice of modelling structure of $\mathbb{E}[U_j(\psi_j)\mid H_j]$, we consider a null (neither the intercept nor covariates are included; i.e., set  $\mathbb{E}[U_j(\psi_j)\mid H_j]$ to zero) model (A1) and a linear model (A3). Specifically, the EEs under A3 for stage 2 and 1 are given by (\ref{A_stage2_example}) and (\ref{A_stage1_example})  where setting $D_{ji}^{\top}\xi_j$ to be zero for $j=1,2$ leads to EEs under A1. As can be seen from Table \ref{tab:sim1_res}, A3 greatly improves the efficiency of the estimator, although an incorrect linear model is assumed for $\mathbb{E}[S_j(\psi_j)|H_j]$. Additionally, we also considered A2 and A4, which are respectively similar to A1 and A3 in terms of modelling, but different in estimation in the sense that the OLS regression is used with the estimated propensity score $\mathbb{E}\left(A_{j}|H_{j},\hat{\alpha}_j\right)$ (a one-dimensional summary of the muti-dimensional $H_j$) included as a covariate to remove the effect of confounding \citep{robins2008estimation,rosenbaum1983central}. This strategy is referred to as regression adjustment for the propensity score \citep{d1998propensity,vansteelandt2014regression}. Specifically, A2 is the OLS regression of $\widehat{V}^{A}_{j+1}$ on $A_jR_j$ and $\mathbb{E}\left(A_{j}|H_{j},\hat{\alpha}_j\right)R_j$; and A4 is the OLS regression of $\widehat{V}^{A}_{j+1}$ on $A_jR_j$, $\mathbb{E}\left(A_{j}|H_{j},\hat{\alpha}_j\right)R_j$ and $D_j$. A3 and A4, compared with A1 and A2, provide doubly robust estimates of $\psi^{\ast}$. Moreover, when $\mathbb{E}[\widehat{V}^{A}_{j+1}|H_j,A_j]$ can be well approximated by a linear combination of $A_jR_j$, $\mathbb{E}\left(A_{j}|H_{j},\hat{\alpha}_j\right)R_j$ and $D_j$, it is expected that A4 will provide a more efficient estimate of $\psi^{\ast}$ than A3. It is worth noting that the unbiased OLS estimators of $\psi^{\ast}$ given by A2 and A4 do not necessitate the correct specification of the relationship between $\mathbb{E}\left(A_{j}|H_{j},\hat{\alpha}_j\right)$ and the conditional mean of  $\widehat{V}^{A}_{j+1}$, which does not generally hold in nonlinear (with a non-identity link function) outcome models \citep{vansteelandt2014regression}. 

Dynamic weighted OLS (dWOLS) developed by \cite{wallace2015doubly} is another approach to estimate $\psi^{\ast}$ in a regression manner where a treatment-based weight is used to account for confounding. In fact, dWOLS is closely related to the g-estimation and we illustrate this connection in the supplementary material. Briefly,  these three approaches, dWOLS, A3 and A4, reported in Table \ref{tab:sim1_res}, are the same in terms of the model specification (the blip functions and working models for $\mathbb{E}[A_j|H_j]$ and $\mathbb{E}[S_j(\psi_j)|H_j]$) and yield consistent estimators of $\psi^{\ast}$ with different standard errors since different estimation methods are used. For IPWE and AIPWE, we specify the class of candidate regimes as $\mathcal{D}_{\tilde{\psi}}=\left\{\left(d_{\tilde{\psi}_1},d_{\tilde{\psi}_2}\right):\left(I\{L_1<\tilde{\psi}_1\},I\{L_2<\tilde{\psi}_2\}\right) \text{for}~ \tilde{\psi}\in R_{+}^2\right\}$ based on a priori knowledge that a patient should be treated when the CD4 count is lower than some critical value, here, $\tilde{\psi}_j$. Obviously, the true optimal treatment regime $d^{\text{opt}}$ is contained in $\mathcal{D}_{\tilde{\psi}}$ with $\tilde{\psi}_1^{\ast}=-\psi_{10}^{\ast}/\psi_{11}^{\ast}=250$ and $\tilde{\psi}_2^{\ast}=-\psi_{20}^{\ast}/\psi_{21}^{\ast}=360$ as the true critical values. When performing IPWE, $\hat{M}_{\psi,2}(\bar{L}_{2i})$ in (\ref{IPWE}) is required, which can be expressed in terms of $\mathbb{E}\left(A_{ji}|H_{ji},\hat{\alpha}_j\right)$ for $j=1,2$. As for $\hat{\omega}_{\psi,j}(\bar{L}_{ij})$ involved in AIPWE in (\ref{AIPWE}), we approximate it by the predicted value of the Q-function on $d_{\tilde{\psi}_j}$, i.e., $Q_j(H_j,d_{\tilde{\psi}_j};\hat{\psi}_j,\hat{\xi}_j)$ with $Q_j(\cdot)$ given in (\ref{Q1_example}) and (\ref{Q2_example}) for $j=1,2$.  AIPWE, similar to A3, A4 and dWOLS, is double-robust and outperforms IPW  as shown in Table \ref{tab:sim1_res}, even though the augmentation term is misspecified. To compare the performance of (A)IPWE and the indirect methods, we present the estimates of $\tilde{\psi}^{\ast}_j, j=1,2$ from the indirect methods in  Table \ref{tab:sim1_res}. The increased robustness of (A)IPWE comes with higher variability than the indirect methods.

\begin{figure}
\centering
\includegraphics[height=7cm,width=12cm]{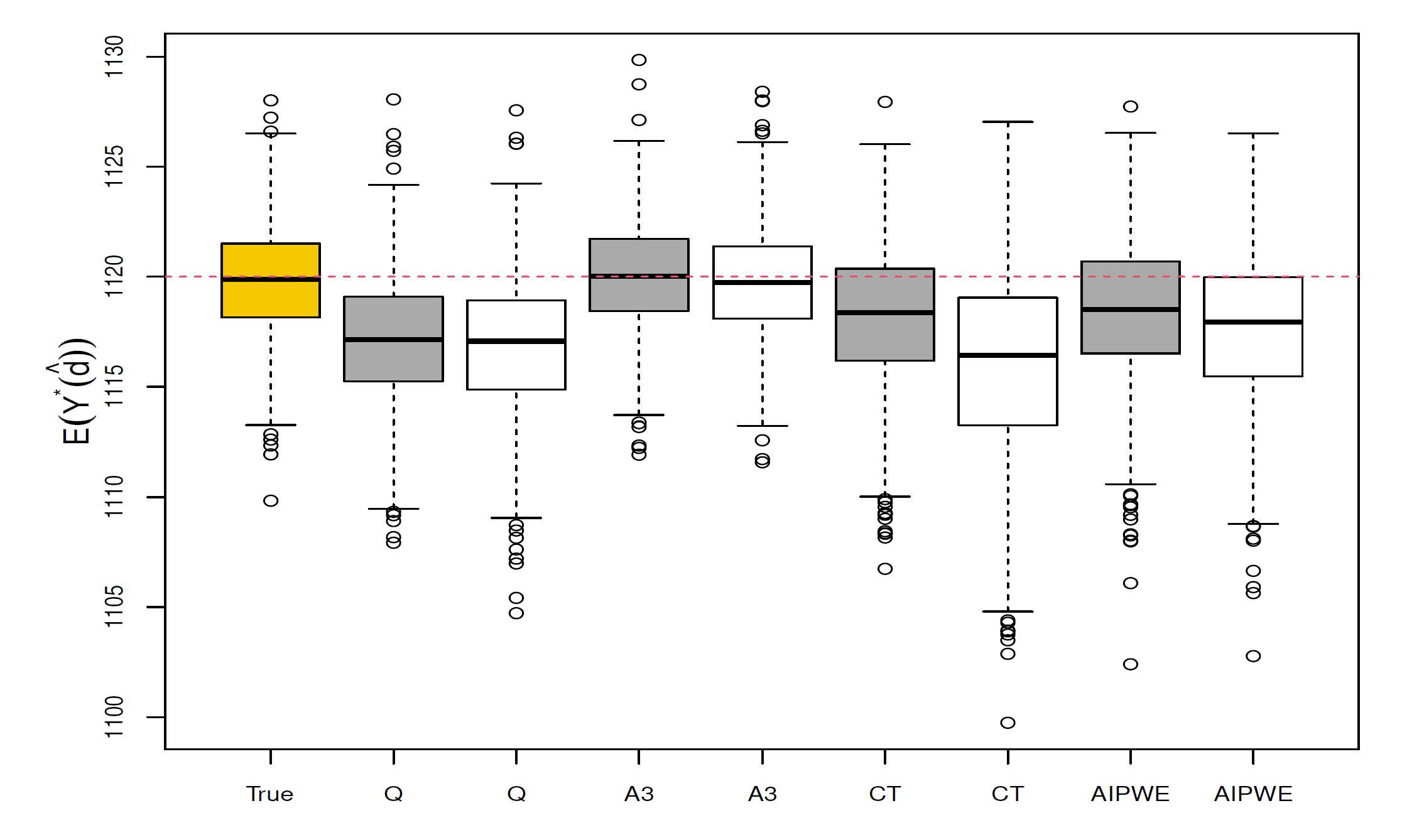}
\caption{\label{fig:marginal_mean} Evaluation of $\mathbb{E}[Y^{\ast}(\hat{d}^{\text{opt}})]$ in case 1 via MC sampling ($B=10000$) with $\hat{d}^{\text{opt}}$ obtained from different methods. For each method, the grey (white) plot corresponds to $\hat{d}^{\text{opt}}$ estimated under the sample size $n=1000$ ($n=500$).}
\end{figure}

In addition to the performance in estimating $\psi^{\ast}$, we also evaluate the different approaches in terms of $\mathbb{E}[Y^{\ast}(\hat{d}^{\text{opt}})]$ (Figure \ref{fig:marginal_mean}) and the decision accuracy (Table \ref{tab:sim1_accu}), which can be used to assess a non-parametric method, for example, causal tree. Here and thereafter, we choose A3 as a representative of A-learning. $\mathbb{E}[Y^{\ast}(\hat{d}^{\text{opt}})]$ is the average value of the outcome that would be achieved if $\hat{d}^{\text{opt}}$ had been followed by the population. For a given $\hat{d}^{\text{opt}}$, $\mathbb{E}[Y^{\ast}(\hat{d}^{\text{opt}})]$ is calculated via Monte Carlo (MC) sampling: $B$ observations complying with $\hat{d}^{\text{opt}}$ are generated ($L_{1b}$, then $L_{2b}|L_{1b},\hat{d}^{\text{opt}}_{1}$ and finally $Y_{b}|\bar{L}_{2b},\hat{d}^{\text{opt}}$ for $b=1,\dots,B$) and $\mathbb{E}[Y^{\ast}(\hat{d}^{\text{opt}})]$ is calculated as the sample mean of $Y$ over these $B$ observations.  The box plots of $\mathbb{E}[Y^{\ast}(\hat{d}^{\text{opt}})]$ in Figure \ref{fig:marginal_mean} are based on 1000 MC replications with $\hat{d}^{\text{opt}}$ obtained from different methods and under different sample sizes.  The variation of $\mathbb{E}[Y^{\ast}(\hat{d}^{\text{opt}})]$ under the true regime, as indicated in the first box plot in  Figure \ref{fig:marginal_mean}, comes from the MC sampling approximation. The decision accuracy refers to the proportion of individuals in the testing set for whom the predicted treatment rule learned from the training set matches the true rule. We calculate the decision accuracy for each of the separate stages, denoted as accu1 and accu2 in Table \ref{tab:sim1_accu}, as well as the accuracy over all stages, i.e., accu in Table \ref{tab:sim1_accu}, under different training sample sizes $n_{\text{tr}}=1000$ and $500$ with the testing sample size fixed to be 1000.  As shown in Table \ref{tab:sim1_accu} and Figure \ref{fig:marginal_mean}, A3 performs the best under both 1000 and 500 sample size cases; causal tree, compared with other parametric methods, is more sensitive to the sample size as expected.  The biased estimate of $\psi^{\ast}$, for example, $\hat{\psi}$ from Q-learning, does not necessarily result in severe loss in $\mathbb{E}[Y^{\ast}(\hat{d}^{\text{opt}})]$ as shown in Figure \ref{fig:marginal_mean}, since in our setting a wrong treatment decision produces a small regret when $L_j$ is close to its critical value, which is implied by the form of the true regret functions. Moreover, the distributions of $L_j$'s also make an impact on the performance of each method in terms of the accuracy and the achieved average value.

\begin{table}
{\small
\centering
\caption{\small Estimation results in case 1. SD is the standard deviation of the estimates across 1000 MC replications; SE is the estimated standard error which is obtained via sandwich variance estimator for A1-A4, and via bootstrap for the others.} \vspace{1mm}
\begin{tabular}{ccrrrrrr}
  \hline
  & &\multicolumn{3}{c}{$n=1000$} &\multicolumn{3}{c}{$n=500$}\vspace{1mm}\\
 &True & \multicolumn{1}{c}{$\hat{\psi}$} & \multicolumn{1}{c}{SD}& \multicolumn{1}{c}{SE}
  & \multicolumn{1}{c}{$\hat{\psi}$} & \multicolumn{1}{c}{SD} &\multicolumn{1}{c}{SE}\\ 
  \hline \\
   A-learning & & & & &\\
  \multirow{6}{*}{A1} &$\psi_{ 10 }^{\ast}=250$ & 255.73  & 81.72& 79.30  &258.54 & 117.54 & 110.43\\ 
  &$\psi_{ 11 }^{\ast}=-1$  &$-1.0140$& 0.1918  & 0.1860 & $-1.0212$ & 0.2759& 0.2591 \\ 
  &$\tilde{\psi}_{1}^{\ast}=250$ &  245.82 & 35.54 & - &238.74 & 59.223 &-\\
  &$\psi_{ 20 }^{\ast}=720$ & 727.81  & 116.39  & 113.33 & 734.95 & 171.25 & 158.80\\  
  &$\psi_{ 21 }^{\ast}=-2$  &$-2.0182$& 0.2262  & 0.2205 & $-2.0343$ & 0.3364 & 0.3095\\ 
  &$\tilde{\psi}_{2}^{\ast}=360$ &  358.71 & 18.24 & -& 357.33 & 26.46 &-\vspace{3mm}\\
  
  %&$E[Y^{op}]=1120$  & 1118.98& 3.5097 &- & 1118.19 & 3.8977 &-\\
\multirow{6}{*}{A2} &$\psi_{ 10 }^{\ast}=250$ & 255.38 & 99.28 & 96.85& 267.11 & 142.33 &133.81\\ 
  &$\psi_{ 11 }^{\ast}=-1$  &$-1.0137$& 0.2335 & 0.2278 & $-1.0406$ & 0.3360& 0.3162 \\
&$\tilde{\psi}_{1}^{\ast}=250$ &  240.91 & 60.03 & - &204.57 &  802.03 &-\\
 &$\psi_{ 20 }^{\ast}=720$  & 731.33 &119.62& 111.05& 738.76& 158.24& 149.14\\
 &$\psi_{ 21 }^{\ast}=-2$   & $-2.0242$ & 0.2318 &0.2156 &$-2.0392$ &0.3076& 0.2912 \\ 
 &$\tilde{\psi}_{2}^{\ast}=360$ &  359.34 & 21.35 & - &358.72 & 30.97 & -\vspace{2mm}\\
 
 %&$E[Y^{op}]=1120$   & 1118.69 & 3.6063 &- & 1117.58 & 4.3281 &-\\
 \multirow{6}{*}{A3}&$\psi_{ 10 }^{\ast}=250$  & 249.22& 17.55 &18.42  & 247.67 & 26.35&  27.20\\ 
 &$\psi_{ 11 }^{\ast}=-1$   & $-0.9986$  & 0.0389  & 0.0406 & $-0.9949$ & 0.0578 & 0.0597\\
 &$\tilde{\psi}_{1}^{\ast}=250$ &  249.29 & 8.5362  & - &248.26 & 13.05 &- \\
 &$\psi_{ 20 }^{\ast}=720$  & 718.96 & 48.36 & 46.91 & 719.00 & 68.30 &66.20\\ 
 &$\psi_{ 21 }^{\ast}=-2$   & $-1.9989$ & 0.0847 & 0.0820& $-1.9992$ & 0.1199 &0.1100\\
  &$\tilde{\psi}_{2}^{\ast}=360$ &  359.32 & 9.4040& - & 358.92 & 13.27 &-\vspace{2mm}\\

 %&$E[Y^{op}]=1120$   & 1119.75 & 3.4016 &- & 1119.67 & 3.4250 &-\\
\multirow{6}{*}{A4} &$\psi_{ 10 }^{\ast}=250$  & 249.83 & 14.46 &14.73 &249.95 & 21.71 & 20.91\\ 
 &$\psi_{ 11 }^{\ast}=-1$   & $-1.0000$ & 0.0332  &0.0334 & $-1.0000$ & 0.0493 & 0.0473\\
&$\tilde{\psi}_{1}^{\ast}=250$ & 249.65 & 6.9486  & - &249.53 & 10.51 &-\\
 &$\psi_{ 20 }^{\ast}=720$  & 719.46 & 19.24  &18.51 & 721.25 & 26.21 & 25.42\\ 
 &$\psi_{ 21 }^{\ast}=-2$  & $-1.9995$ & 0.0358 &0.0340 & $-2.0029$ & 0.0481 & 0.0467 \\
&$\tilde{\psi}_{2}^{\ast}=360$ &  359.77 & 3.8788 & - &360.01 &5.4653 &-\vspace{2mm}\\

 %&$E[Y^{op}]=1120$ & 1119.80 & 3.3851 &- & 1119.85 & 3.3916 &-\\
 
 \multirow{6}{*}{dWOLS} &$\psi_{ 10 }^{\ast}=250$  & 248.48 &  17.13 &18.55 &246.17 & 28.05 &26.16\\ 
 &$\psi_{ 11 }^{\ast}=-1$   & $-0.9966$ &  0.0388  &0.0408 & $-0.9915$ & 0.0618 &0.0570\\ 
 &$\tilde{\psi}_{1}^{\ast}=250$ & 249.21 & 8.4878  & - &248.11 &  13.10  &-\\
 &$\psi_{ 20 }^{\ast}=720$  & 717.00 & 43.49  & 41.96 & 715.54 & 56.22 &56.60\\ 
 &$\psi_{ 21 }^{\ast}=-2$  & $-1.9942$ & 0.0773 &0.0732 & $-1.9776$ & 0.0984 & 0.0986\\
 &$\tilde{\psi}_{2}^{\ast}=360$ & 359.03 & 8.5790 & - & 358.38 &12.15 &-\vspace{2mm}\\

 %&$E[Y^{op}]=1120$ & - & - &- & - & - &-\\ 
 Q-learning & & & \\
&$\psi_{ 10 }^{\ast}=250$  & 155.49 & 21.76 & 22.73& 154.89 & 32.53 &32.03\\ 
 &$\psi_{ 11 }^{\ast}=-1$ & $-0.7775$ & 0.0491 & 0.0505 &$-0.7757$ & 0.0712&0.0710\\ 
 &$\tilde{\psi}_{1}^{\ast}=250$ &  199.06 & 16.30  & - &197.54 &25.51 &- \\
&$\psi_{ 20 }^{\ast}=720$ & 506.50 & 48.78 &48.32 &508.87 & 65.84&65.75\\ 
&$\psi_{ 21 }^{\ast}=-2$ & $-1.5841$ & 0.0909& 0.0902 & $-1.5879$ & 0.1234&0.1227 \\ 
&$\tilde{\psi}_{2}^{\ast}=360$ &  319.05 & 13.48 & - & 319.22 & 18.45 & -\vspace{2mm}\\

%&$E[Y^{op}]=1120$& 1117.15& 3.6394 & & 1116.96 & 3.9868 & \\
 \multirow{2}{*}{IPWE}& & & \\
&$\tilde{\psi}_{1}^{\ast}=250$ & 260.19 &79.41 & 78.61 & 283.53 & 94.10 &92.07 \\
&$\tilde{\psi}_{2}^{\ast}=360$ & 390.41 &62.49 &61.46 & 398.76 & 73.29 &73.54 \\
AIPWE & & & \\
&$\tilde{\psi}_{1}^{\ast}=250$ & 239.94 &60.79 & 57.84 & 223.67 & 79.82 &75.46 \\
&$\tilde{\psi}_{2}^{\ast}=360$ & 362.30 &20.02 &19.70 & 365.14 & 24.75 &24.70
\\
\hline
\end{tabular}
\label{tab:sim1_res}}
\end{table}

\begin{table}
\centering
\caption{Decision accuracy (SD) [\%]  on the testing set (the sample size in the testing  set is fixed to be $n_{\text{tes}}=1000$) in case 1.} 
\begin{tabular}{crrrrrr}
  \hline\\
  & \multicolumn{3}{c}{$n_{\text{tr}}=1000$} &\multicolumn{3}{c}{$n_{\text{tr}}=500$}\vspace{1mm}\\
 & \multicolumn{1}{c}{accu1} & \multicolumn{1}{c}{accu2}& \multicolumn{1}{c}{accu}
  & \multicolumn{1}{c}{accu1}& \multicolumn{1}{c}{accu2}&\multicolumn{1}{c}{accu}\\
  \hline
Q & 95.60 (1.36)  &95.73 (1.32)&  92.06 (1.64) &  95.13 (1.71) &94.84 (1.80) &91.24 (2.17)\\
A3 & 99.17 (0.70) & 99.15 (0.65) & 98.35 (0.84) & 98.81 (0.94) &98.81 (0.99) & 97.69 (1.32)  \\
CT & 96.27 (2.89) & 97.91 (1.61) & 94.45 (2.92)&93.24 (3.83) &96.00 (2.63) & 89.88 (4.38) 
\\ 
AIPWE & 95.99 (3.10)&  97.80 (1.68)&  94.04 (3.27)& 95.64 (3.18) & 97.82 (1.78) & 93.73 (3.53)\\
\hline
\end{tabular}
\label{tab:sim1_accu}
\end{table}

\subsection{Three decision points with two-dimensional $L_j$}\label{sec:simcase2}
\begin{figure}
\centering
\includegraphics[height=4cm,width=15cm]{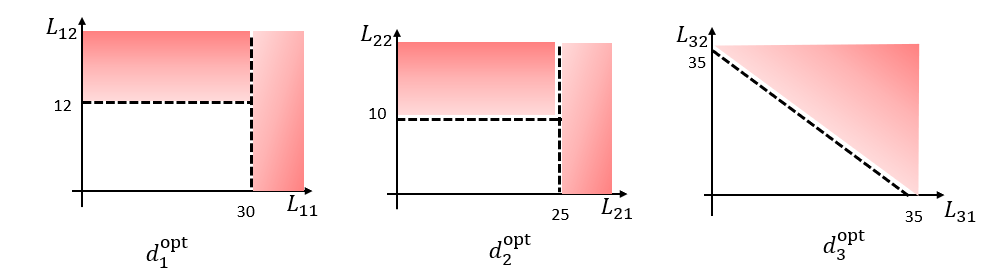}
\caption{\label{fig:case2_true} An illustration of the true treatment rule $d_j^{\text{opt}}(L_{j1},L_{j2})$ for $j=1,2,3$ in case 2. The red areas indicate cases where the treatment should be assigned according to the true rules.}
\end{figure}

Let $N_{(a,b)}(\mu,\sigma^2)$ denotes the truncated normal distribution constrained in the interval $(a,b)$. The observed data in this section are generated as follows: $W \sim N_{(10,\infty)}(45,10^2)$, $L_{11}\sim N_{(0,\infty)}(20,5^2)$, $L_{12}\sim N_{(0,\infty)}(10,3^2)$,
$A_1|H_1\sim \text{Bin}(1,p_1)$ with $p_1=\text{expit}(-3+0.1W)$; 
$L_{21}|H_1,A_1\sim N_{(0,\infty)}(1.25L_{11}-2A_1,5^2)$, $L_{22}|H_1,A_1\sim N_{(0,\infty)}(L_{12}-A_1,3^2)$,
$A_2|H_2\sim \text{Bin}(1,p_2)$ with $p_2=\text{expit}(-1+0.04(L_{21}+L_{22}))$; 
$L_{31}|H_2,A_2\sim N_{(0,\infty)}(L_{21}-2(A_1+A2),5^2)$, $L_{32}|H_2,A_2\sim N_{(0,\infty)}(L_{22}-A_2,3^2)$,
$A_3|H_3\sim \text{Bin}(1,p_3)$ with $p_3=\text{expit}(-2+0.1L_{31})$; and 
$$Y|H_3,A_3 \sim  N(100,10^2)-\mu_1(H_1,A_1)-\mu_2(H_2,A_2)-\mu_3(H_3,A_3),$$
where 
$\mu_1(H_1,A_1)=0.5(|L_{11}-30|+|L_{12}-12|)\times(I\{L_{11}>30 ~\text{or} ~L_{12}>12\}-A_1)^2$,
$\mu_2(H_2,A_2)=(|L_{21}-25|+|L_{22}-10|)\times(I\{L_{21}>25 ~\text{or} ~L_{22}>10\}-A_2)^2$ and
$\mu_3(H_3,A_3)=2\log(W)\times(I\{L_{31}+L_{32}>35\}-A_3)^2$. See Figure \ref{fig:case2_true} for an illustration of the pattern of the true treatment rule in each stage. Briefly, $d_{j}^{\text{opt}}$ has a boolean structure with respect to biomarkers $L_{j1}$ and $L_{j2}$ for $j=1,2$; and $d_{3}^{\text{opt}}$ takes a linear form.

\begin{figure}
\centering
\includegraphics[height=9cm,width=13cm]{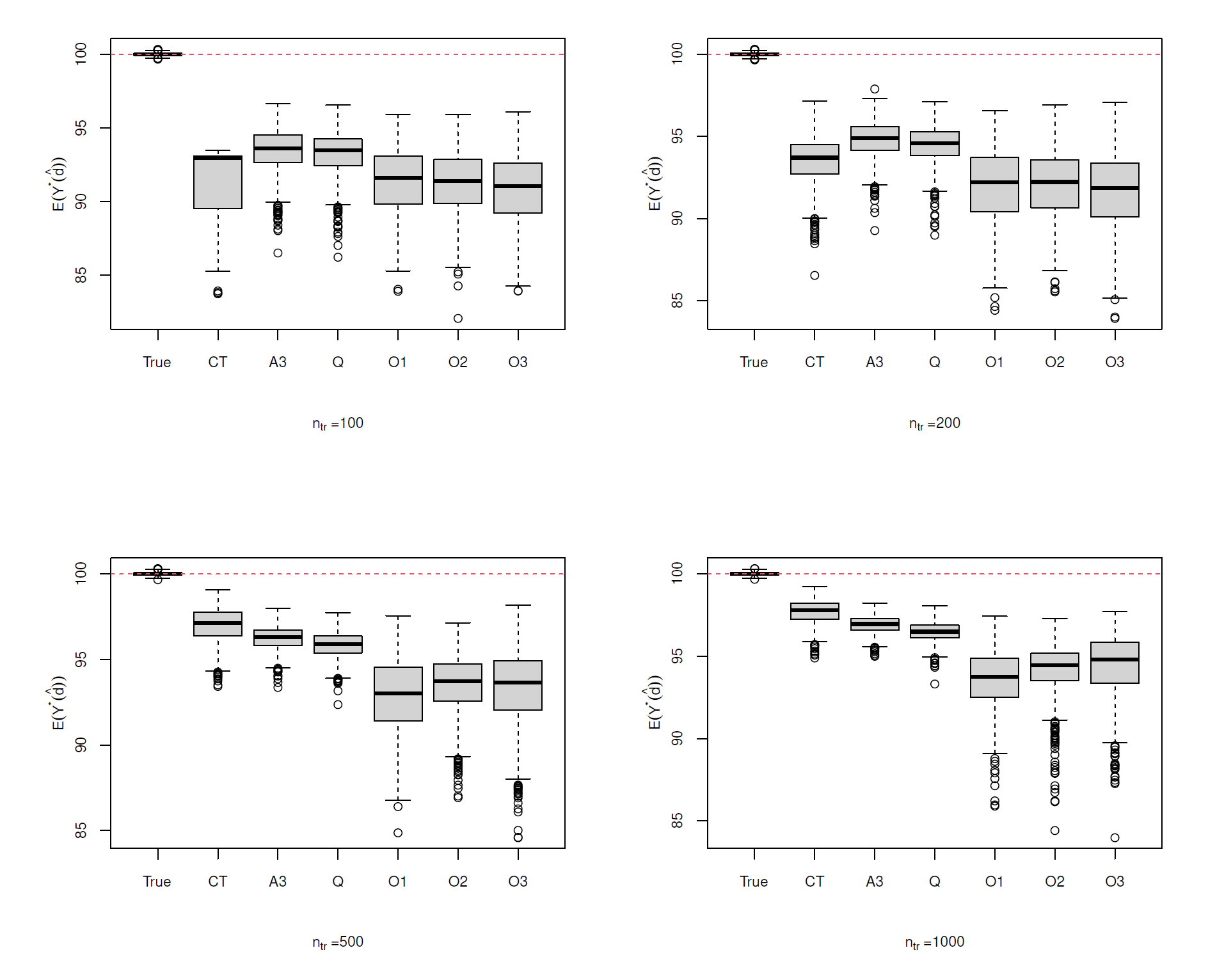}
\caption{\label{fig:marginal_mean_case2} Evaluation of $\mathbb{E}[Y^{\ast}(\hat{d}^{\text{opt}})]$ in case 2 via MC sampling ($B=10000$) with $\hat{d}^{\text{opt}}$ obtained from different methods under different training sample sizes.}
\end{figure}

In addition to the methods performed in Section \ref{sec:simcase1},  OWL and its variants are also implemented for this complicated setting. Specifically, we perform BOWL with linear decision functions for all stages (called O1), AOL with linear decision functions (called O2) and AOL with decision functions determined by radial basis function (RBF) kernel (called O3). The implementation of O1, O2 and O3 uses R package \texttt{DTRlearn2} with tuning parameters chosen by four-fold cross-validation (default setting). For indirect methods, we still specify a linear model for the Q-function in Q-learning and a linear model for the blip function in A-learning while modelling $\mathbb{E}[A_j\mid H_j]$ correctly. Besides Q-learning, A-learning also suffers from model misspecification even in stage 3. This is because a linear blip function $\gamma_3(H_3,A_3;\psi_3)=A_3R_3^{\top}\psi_3$ will lead to a regret function with the regret scale as $|R_3^{\top}\psi_3|$, which is obviously different from that in $\mu_3(H_3,A_3)$. It is thus expected that CT,  AIPWE and OWL will work better than Q- and A-learning. However, as AIPWE in this case involves optimizing a discontinuous objective function over at least six parameters, it produces very unreliable results in our simulation. We therefore only compare the performance of Q-learning, A-learning, CT and OWL (O1, O2 and O3) in terms of the decision accuracy (Table \ref{tab:sim2_accu}) and the value $\mathbb{E}[Y^{\ast}(\hat{d}^{\text{opt}})]$ (Figure \ref{fig:marginal_mean_case2}) under different training sample sizes. The performance of CT, as shown in Table \ref{tab:sim2_accu} and Figure \ref{fig:marginal_mean_case2}, is sensitive to the sample size and when $n_{tr}$ gets sufficiently large  e.g., $n_{tr}=500$ and 1000, CT outperforms the other methods, especially in terms of `accu1' and `accu2' in Table \ref{tab:sim2_accu} as the true decision rules in stage 1 and 2 exhibit boolean structures. OWL methods, though being robust to model misspecification, do not perform better than the others in this setting. The main reason is that the biomarker information $L_j$ is of low dimension, in which case the regression-based indirect methods perform pretty well especially in large sample sizes whereas the decision functions in OWL are dominated by the support points. OWL methods are desirable options when the sample size is small or when the covariate information is high-dimensional as illustrated in the simulation studies in \cite{zhao2012estimating,zhao2015new}

\begin{table}
\centering
\caption{Decision accuracy (SD) [\%] on the testing set ($n_{\text{tes}}$  fixed to be 1000) in case 2 given by different methods under different training sample sizes. O1 stands for BOWL with linear decision functions; O2 (O3) stands for AOL with linear (nonlinear, specifically using RBF kernel) decision functions.} 
\small{
\addtolength{\tabcolsep}{-0.2em}
\begin{tabular}{crrrrrrrrr}
  \hline \vspace{-2mm} \\
 & \multicolumn{1}{c}{\textbf{accu1}} & \multicolumn{1}{c}{\textbf{accu2}} & \multicolumn{1}{c}{\textbf{accu3}} &\multicolumn{1}{c}{\textbf{accu}} & \multicolumn{1}{c}{\textbf{accu1}} & \multicolumn{1}{c}{\textbf{accu2}} & \multicolumn{1}{c}{\textbf{accu3}}&\multicolumn{1}{c}{\textbf{accu}}
\\ \cmidrule(rl){2-5} \cmidrule(rl){6-9} \vspace{-1.5mm} \\
& \multicolumn{4}{c}{$n_{\text{tr}}=100$} &\multicolumn{4}{c}{$n_{\text{tr}}=200$}\vspace{1mm}\\
Q &74.13 (12.0)&73.29 (7.40) & 67.79 (8.94) & 36.65 (8.83) 
&79.83 (8.60)& 78.31 (4.87)& 73.52 (7.38)& 45.65 (7.51)\\

A3 &73.39 (12.3) &74.08 (6.83)&70.64 (8.63) & 38.34 (8.92)
& 79.05 (8.52)&79.10 (4.19)& 77.37 (6.75)& 48.21 (7.41)\\ \vspace{1mm}

CT &66.32 (16.3) &62.02 (11.9)&63.01 (14.3) & 22.28 (6.64) 
& 70.09 (11.5) & 72.99 (6.19) &73.03 (11.4) &38.01 (9.77)\\

O1 &62.83 (18.3) &63.52 (13.4) & 62.32 (14.3) & 25.36 (10.9) 
&64.41 (18.1) & 65.65 (13.4) & 67.03 (13.2)& 28.71 (11.3)\\
O2 &61.61 (18.0) &63.80 (12.4) & 61.84 (15.0) & 24.60 (10.1) 
&63.85 (18.6)& 68.09 (11.0)& 67.34 (13.4) &30.02 (11.5)\\
O3 &59.29 (16.8) &61.89 (14.0) & 61.17 (15.8) & 22.63 (10.1)
&60.88 (16.4)&66.09 (14.0) & 66.43 (15.3) & 27.13 (11.6)  \\ \vspace{1mm}\\
& \multicolumn{4}{c}{$n_{\text{tr}}=500$} &\multicolumn{4}{c}{$n_{\text{tr}}=1000$}\vspace{1mm}\\ 
Q & 84.89 (5.73)& 82.48 (2.46) & 80.11 (5.15) & 55.59 (5.60)
&  87.20 (4.39)& 84.03 (1.53) & 83.48 (4.22) &  60.59 (4.79)  \\

A3 & 84.39 (5.91) & 82.94 (2.11) & 84.70 (4.15) & 58.95 (5.40)
&  87.15 (4.47) & 84.30 (1.43) & 88.81 (3.10) & 64.87 (4.35) \\ \vspace{1mm}

CT & 85.25 (12.1) & 92.10 (5.67) &79.04 (7.98) &62.06 (11.3)
& 90.20 (8.19) & 93.57 (4.88) &81.84 (6.62) & 69.10 (9.01) \\  

O1 & 64.92 (18.6)&69.08 (12.9) & 73.84 (11.6) & 33.82 (12.7)
&68.39 (16.8) &72.29 (11.2) &77.01 (11.0) & 38.88 (12.0) \\  
O2 & 70.21 (14.5) &74.64 (7.73) & 73.75 (11.3) & 39.23 (10.7)
&73.72 (12.5) &76.94 (6.66) &76.59 (11.6) &44.14 (10.3) \\
O3 & 66.70 (15.8) &72.91 (12.9) & 74.22 (14.1) & 36.73 (13.4)
&70.66 (16.8) &77.74 (10.8) &79.16 (11.9)& 43.91 (13.9)\\
\\\hline
\end{tabular}}
\label{tab:sim2_accu}
\end{table}

\section{Application to $\text{STAR}^{\ast}\text{D}$} \label{sec:app}
In this section, we apply the aforementioned methods to the analysis of Sequenced Treatment Alternatives to Relieve Depression (STAR*D) data, which is a randomized clinical trial designed to determine the most effective treatment strategies for patients with major depressive disorder who did not achieve remission from initial treatment \citep{rush2004sequenced}. The trial proceeded as follows: all eligible participants were treated with citalopram (CIT) at level 1. Those without sufficient improvement, determined by the clinician based on the 16-item Quick Inventory of Depressive Symptomatology ($\text{QIDS-C}_{16}$, the higher the worse), moved to level 2 and were randomized to one of seven treatment options (four \textit{switch} options: switch from CIT to sertraline, bupropion, venlafaxine or cognitive therapy; three \textit{augmentation} options: augment CIT with bupropion, buspirone or cognitive therapy) at the beginning of level 2 based on their preference between switch and augmentation. Those receiving cognitive therapy (either switch or augmentation) but without sufficient improvement at level 2 then moved to level 2A and were randomized to one of two  switch options (switch to bupropion or venlafaxine) at the beginnning of level 2A. Those without sufficient improvement at level 2 (level 2 or level 2 plus level 2A if cognitive therapy was assigned in level 2) then moved to level 3 and were randomized to one of  four treatment options (two switch options: switch to mirtazapine or nortriptyline; two augmentation options: augment with lithium or thyroid hormone) at the beginning of level 3. Finally those without sufficient improvement at level 3 moved to level 4 with random assignment to one of two switch options (tranylcypromine or the combination of mirtazapine and venlafaxine). At each treatment level, clinical visits were scheduled at weeks 0, 2, 4, 6, 9 and 12 until sufficient improvement was achieved. Moreover, participants with intolerance of current treatment or minimal reduction in $\text{QIDS-C}_{16}$ were encouraged to move to the next treatment level earlier. Those achieving partial (not sufficient) improvement at week 12 tended to continue with current treatment level for two more weeks to identify whether sufficient improvement could be achieved. This resulted in different duration times in each level between different participants. Sufficient improvement is defined as $\text{QIDS-C}_{16}\leq 5$ for at least two weeks and at the same time without intolerable side effects. Participants with sufficient improvement at that level stopped from moving to the next level and instead moved to a naturalistic follow-up phase. 

\begin{figure}
\centering
\includegraphics[height=5cm,width=10cm]{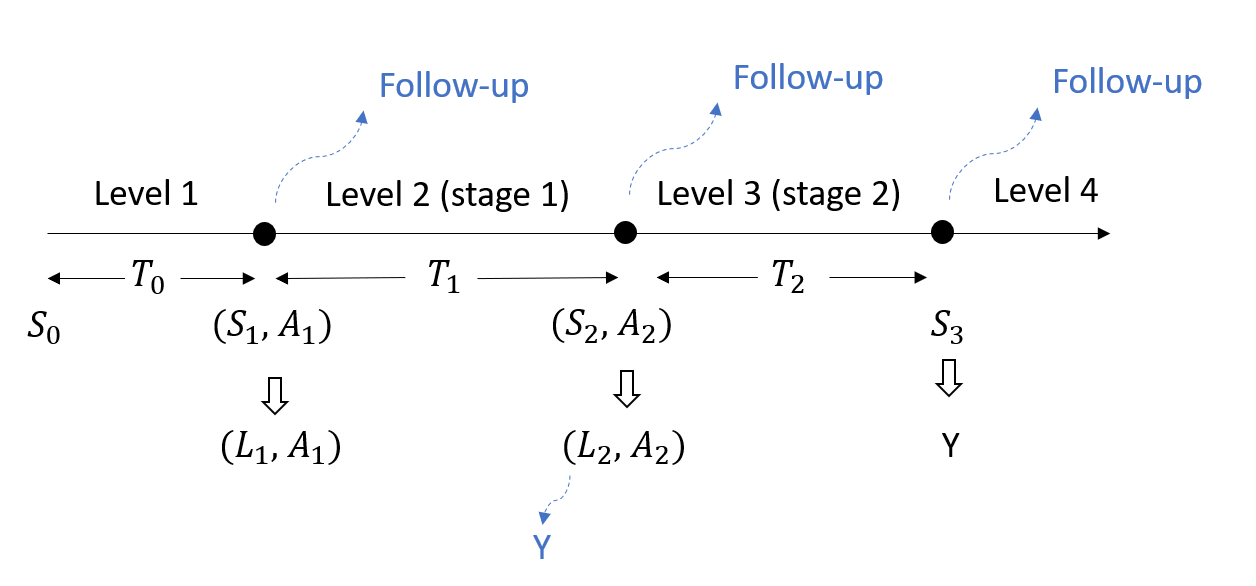}
\caption{\label{fig:stard} Illustration of STAR*D design.}
\end{figure}

Following \cite{schulte2014q}, we restrict our attention to the effect of the two main treatment strategies, i.e., switch versus augmentation of the current treatment, and ignore the substrategies within switch or augmentation. Taking level 2A as a part of level 2, we focus on level 2 and level 3 and define them as stage 1 and stage 2 to adapt to the framework described in Section \ref{sec:model}. Let $A_j, j=1, 2$ denote the treatment received in stage $j$ with $A_j=1$ for switch and $A_j=0$ for augmentation. Let $S_j,j=1, 2$ denote the $\text{QIDS-C}_{16}$ score at the beginning of stage $j$ (equivalently, at the end of stage $j-1$), which was collected prior to the the treatment decision $A_j$, and $S_3$ denotes the $\text{QIDS-C}_{16}$ score at the end of stage 2 (level 3). Define $\Delta S_{j}=(S_j-S_{j-1})/T_{j-1}, j=1, 2$ as the slope of $\text{QIDS-C}_{16}$ score over stage $j-1$ (level $j$)  with $S_0$ denoting the baseline $\text{QIDS-C}_{16}$ score obtained at study entry and $T_{j-1}$ being the time spent in stage $j-1$ (level $j$).  Define $L_1=(S_0,S_1,\Delta S_{1},T_{0})$ as the covariate information prior to $A_1$ and let $L_2=(S_2,\Delta S_{2},T_{1},I_{\text{sim}})$ denote the covariate information prior to $A_2$ with $I_{\text{sim}}$ being the indicator of whether sufficient improvement was achieved at the end of stage 1 (level 2). Define the final outcome $Y$ to be the average of negative $\text{QIDS-C}_{16}$ scores at the end of available stages, i.e, $Y=-(1-I_{\text{sim}})(S_2+S_3)/2-I_{\text{sim}}S_2$.  Note that there were participants dropping out of the study when moving from level 2 to level 3, i.e., who refused to enter level 3 though no sufficient improvement was achieved at level 2. After deleting the dropouts, we have 815 participants in stage 1 (level 2):  329 of them moved to stage 2 (level 3), for whom the observed records are $(L_1,A_1,L_2,A_2,Y)$; and the others (486 participants with sufficient improvement at level 2) entered the follow-up phase after level 2, for whom the observed records are $(L_1,A_1,Y)$(see Figure \ref{fig:stard} for an illustration). Therefore, when using methods which perform through backward induction as shown in Algorithm \ref{algorithm}, we only need to specify how $S_3$ depends on the history in the final stage (stage 2).

In Q-learning, the Q-function in stage 2 (level 3) is $Q_2(\bar{L}_2,\bar{A}_2;\psi_2,\xi_2)=-I_{\text{sim}}S_2-(1-I_{\text{sim}})(S_2+\mathbb{E}[S_3|\bar{L}_2,\bar{A}_2,I_{\text{sim}}=0;\psi_2,\xi_2])/2$, where $\mathbb{E}[S_3|\bar{L}_2,\bar{A}_2,I_{\text{sim}}=0;\psi_2,\xi_2]=A_2(\psi_{20}+\psi_{21}\Delta S_{2})+D_2^{\top}\xi_2$ with $D_2=(1,S_2,S_1,S_0)$. That is, the treatment effect of $A_2$ is allowed to depend on the slope of $\text{QIDS-C}_{16}$ score over stage 1, which is exactly the same as that in \cite{schulte2014q}, and the treatment-free term here is different from that in \cite{schulte2014q} where $D_2$ was set as $(1,S_2,\Delta S_{2})$. In terms of Q-Q plot shown in Figure S1 in the supplement, setting $D_2$ as $(1,S_2,S_1,S_0)$ gives a slightly better mean model for $S_3$. As shown in Table \ref{tab:stard_par} (see the rows corresponding to Q-learning), these two different specifications of the treatment-free term lead to different estimation results of $\psi_2$, especially for $\psi_{21}$. Although the estimates of $\psi_{21}$ are different in terms of scales and signs, they are both shown to be non-significant. Given $\hat{\psi}_2$ and $\hat{\xi}_2$, the value function obtained from Q-learning in stage 2 is therefore $\tilde{V}_2^{Q}=-I_{\text{sim}}S_2-(1-I_{\text{sim}})[S_2+I\{\hat{\psi}_{20}+\hat{\psi}_{21}\Delta S_{2}>0\}(\hat{\psi}_{20}+\hat{\psi}_{21}S_{2s})+D_2^{\top}\hat{\xi}_2]/2$, for which we specify the Q-function in stage 1 as $Q_1(L_1,A_1;\psi_1,\xi_1)=A_1(\psi_{10}+\psi_{11}\Delta S_{1})+D_1^{\top}\xi_1$ with $D_1$ being $(1,S_1,S_0)$ or $(1,S_1,\Delta S_{1})$. The estimation results of $\psi_1$ under two different settings of $D_1$ are given in Table \ref{tab:stard_par}: different in scales especially for $\psi_{11}$ but consistent in the sign and in being non-significant. For A-learning, the propensity scores in stage 2 and stage 1 are specified as $P(A_2=1|\bar{L}_2,I_{\text{sim}}=0,A_1;\alpha_2)=\text{expit}(\alpha_{20}+\alpha_{21}S_2+\alpha_{22}T_1+\alpha_{23}A_1)$ and $P(A_1=1|L_1;\alpha_1)=\text{expit}(\alpha_{10}+\alpha_{11}S_1+\alpha_{12}T_0)$, respectively; which are different from those used in \cite{schulte2014q} where $(S_2,\Delta S_{2},A_1)$ and $(S_1,\Delta S_{1})$ were predictors incorporated in the logistic regression models. See Figure S2 in the supplement for a comparison between these two models. Additional to the models for the propensity scores, we keep the treatment-free terms in A-learning consistent with those in Q-learning, and the estimation results are given in Table \ref{tab:stard_par}.  A-learning produces similar results with those from Q-learning.  However, the results from A-learning appear to be less affected by the specification of the treatment-free terms, especially in stage 1. This can be seen from  Table \ref{tab:stard_par} by comparing the paired black and blue rows under each treatment-free setting.

\begin{figure}
\centering
\includegraphics[height=4.5cm,width=13cm]{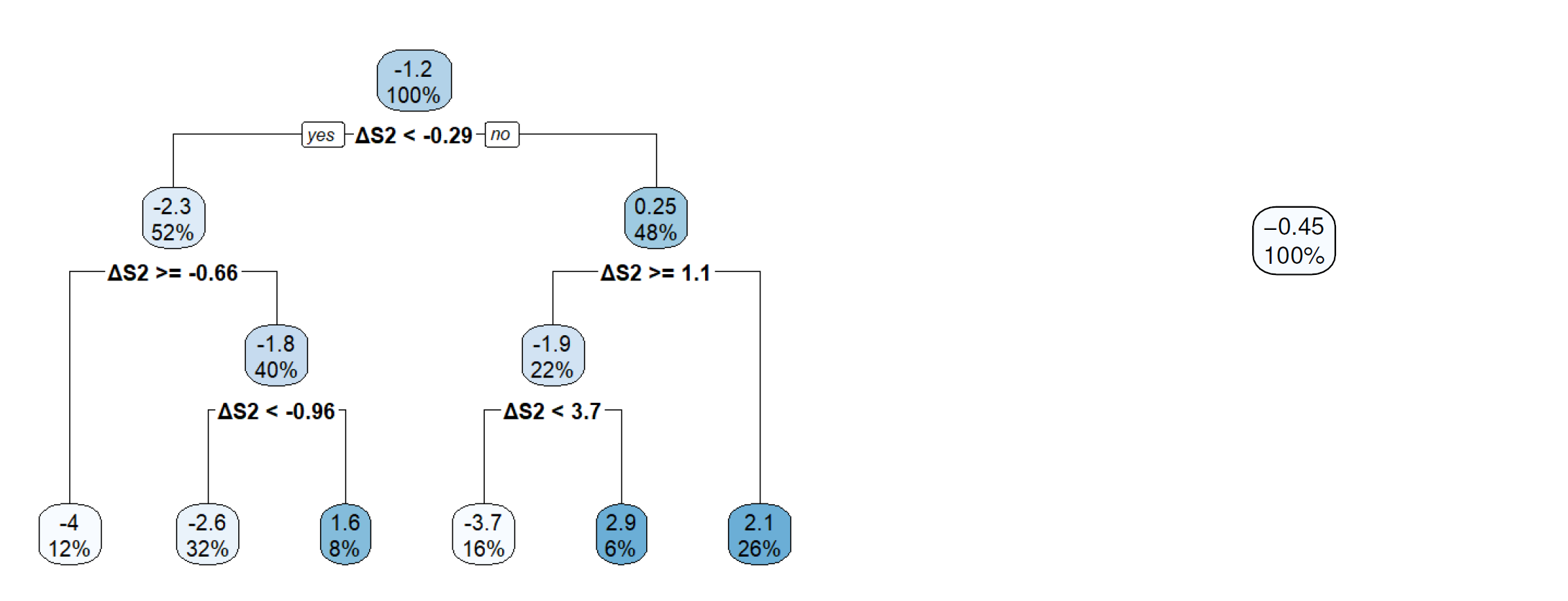}
\caption{\label{fig:stard_tree} Analysis of STAR*D by the causal tree method. Left (Right): the estimated contrast function, indicated by the numbers in the terminal nodes/leaves, in stage 2 (1).}
\end{figure}

We also apply AIPWE to STAR*D, where the class of candidate regimes is $\mathcal{D}_{\psi}=\{d_{\psi}=\left(d_{\psi_1},d_{\psi_2}\right):\left(I\{\psi_{10}+\psi_{11}\Delta S_{1}>0\},I\{\psi_{20}+\psi_{21}\Delta S_{2}>0\}\right) \text{for}~ \psi\in R^4\}$. Note that $d_{\psi}$, expressed via four parameters, is not unique since for any given $(\psi_1,\psi_2)$, $(d_{c_1\psi_1},d_{c_2\psi_2})$ indicates the same treatment regime whatever values of $(c_1,c_2)\in R_{+}^2$. To achieve uniqueness, we follow the strategy in \cite{zhang2012robust} and normalize the obtained parameters from maximizing (\ref{AIPWE}) to ensure $\|\psi_1\|_2=1$ and $\|\psi_2\|_2=1$.  Note that the strategy used in the simulation studies, i.e, forcing the $\psi_{11}$ and $\psi_{21}$ to be $-1$, does not work for STAR*D analysis because we are uncertain as to the direction of the interaction between $A_j$ and $\Delta S_{j}$ for $j=1,2$. The standard errors (via bootstrap) in parentheses in  Table \ref{tab:stard_par} are calculated after normalizing the estimated values of $\psi$. The estimated value of $\psi_{11}$ is positive from AIPWE, whereas  Q- and A-learning produce negative estimates of $\psi_{11}$. This makes it difficult to reach a conclusion on how $\Delta S_{1}$, the slope of $\text{QIDS-C}_{16}$ over level 1, interacts with $A_1$ as the result from AIPWE suggests switching if $\text{QIDS-C}_{16}$ increases rapidly over level 1 ($\Delta S_{1}>2.5$, 2.6\% participants) while the results from  Q- and A-learning suggest switching if $\text{QIDS-C}_{16}$ decreases rapidly over level 1 (Q- and A-learning under different settings give different thresholds, e.g., $\Delta S_{1}<-7.5$, 3.3\% participants; $\Delta S_{1}<-5$, 7.6\% participants), although $\hat{\psi}_{11}$ is found to be non-significant by all these methods. The result from casual tree, as shown in Figure \ref{fig:stard_tree}, suggests a homogeneous effect of $A_1$ (with the treatment in stage 2 fixed to be the optimal one) on the defined final outcome, that is, augmentation is recommended for everyone in stage 1. Additionally, we did not apply OWL methods to STAR*D as attention is restricted to how the effect of $A_j$ varies with one-dimensional $\Delta S_{j}$.

The difference in model assumptions involved in these different methods is the main reason for the difference in the estimation results obtained in Table \ref{tab:stard_par} and Figure \ref{fig:stard_tree}. Additional to models for propensity scores and treatment-free terms, Q- and A-learning impose a linear (in $\Delta S_{j}$) contrast function whereas the causal tree method relaxes this restriction and yields a non-monotonic contrast function as shown in the left of Figure \ref{fig:stard_tree}. The result from the causal tree suggests switching in stage 2 (level 3) when $\Delta S_{2}>3.7$ or $-0.29<\Delta S_{2}<1.1$ or $-0.96<\Delta S_{2}<-0.66$. Participants with $\Delta S_{2}>3.7$ gain the biggest reward from switching. Based on the obtained results and with the aim of reaching a straightforward and interpretable conclusion on treatment regimes, we refit the Q- and A-learning with $\Delta S_{2}$ factorized into $\Delta S_{2}>3.7$ and $\Delta S_{2}\leq3.7$, i.e, the contrast function in stage 2 is specified as $\psi_{20}+\psi_{21}I\{\Delta S_{2}>3.7\}$. We also restrict the effect of $A_1$ to be unrelated to $\Delta S_{1}$,  i.e, the contrast function in stage 1 is specified as $\psi_{10}
$. The propensity scores and the treatment-free term are specified in the same way as before (i.e., $D_2=(1,S_2,S_1,S_0)$ and $D_1=(1,S_1,S_0)$). The results from Q- and A-learning under this setting are given in Table \ref{tab:stard_par_3.7}: participants should augment CIT in stage 1 (level 2) and those with $\Delta S_{2}>3.7$ at the end of stage 1 (level 2) should switch to another treatment in stage 2 (level 3). When applying AIPWE in this setting, the treatment rule in stage 2 indexed by $\psi_{20}$ (with $\psi_{21}$ fixed to be 1), $d_{\psi_{20}}=I\{\psi_{20}+I\{\Delta S_{2}>3.7\}>0\}$, in fact only indicates three possible options (1) switching for all participants if $\psi_{20}>0$, (2) switching for those with $\Delta S_{2}>3.7$ if $-1<\psi_{20}\leq 0$ and (3) augmentation for all participants if
$\psi_{20}\leq -1$; and similarly the treatment rule in stage 1 indexed by $\psi_{10}$, $d_{\psi_{10}}=I\{\psi_{10}>0\}$, indicates two possible rules: (1) switching for all participants if $\psi_{10}>0$ and (2) augmentation for all participants if $\psi_{10}\leq 0$. Therefore, there are overall six dynamic treatment regimes in $\mathcal{D}_{\psi}$, which can be represented by $\mathcal{D}_{\psi}=\{(d_{\psi_{10}},d_{\psi_{20}}): \psi_{10}\in \{-1,1\}, \psi_{20}\in \{-2,-0.5,1\}\}$ and AIPWE selects the one (from six candidates) which maximizes the objective function given by (\ref{AIPWE}). The estimated value  of $(\psi_{10},\psi_{20})$ from AIPWE is $(-1,-0.5)$, suggesting the same treatment regime as that from Q- and A-learning. As for the uncertainty in AIPWE, 155 among 200 (77.5\%) bootstrap samples select  $(\psi_{10},\psi_{20})$ to be $(-1,-0.5)$.

\begin{figure}
\centering
\includegraphics[height=6cm,width=12cm]{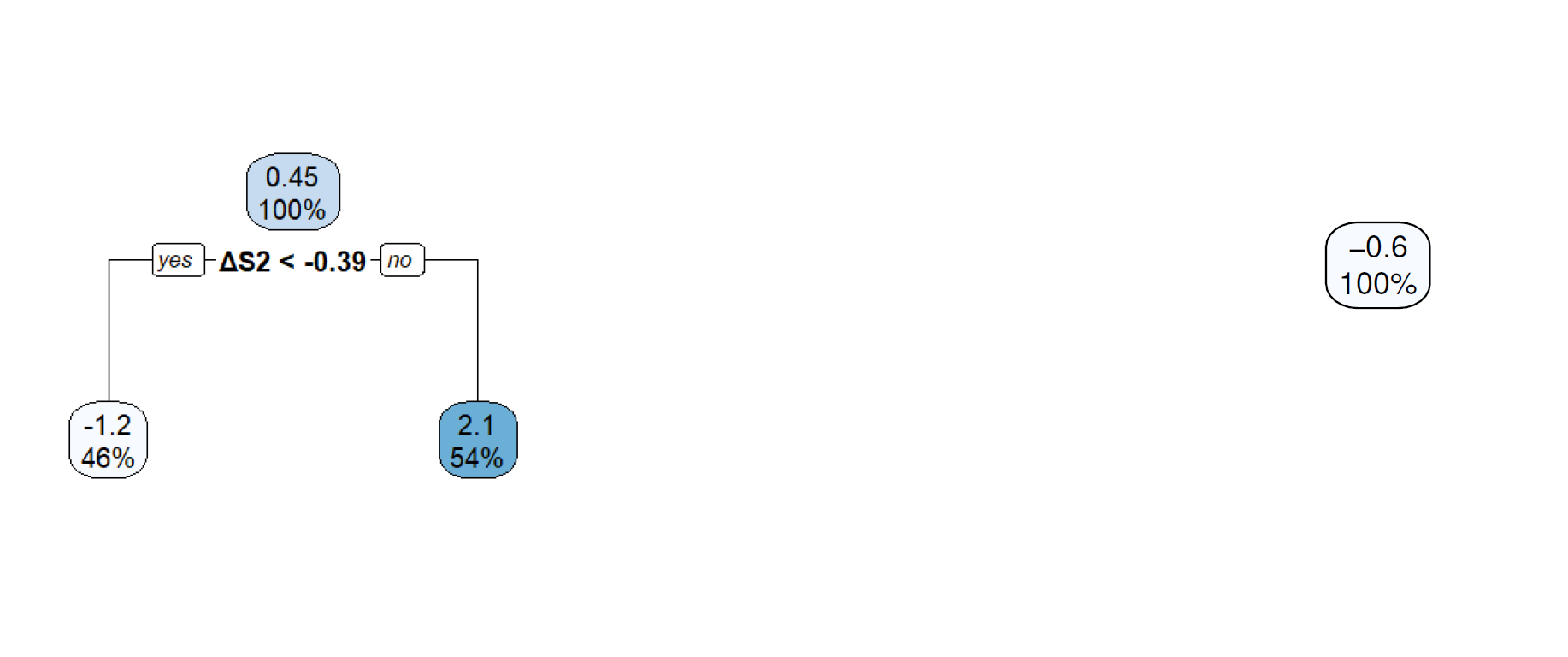}
\caption{\label{fig:stard_tree_hrsd} Results from causal tree method with  $\text{HRSD}_{17}$ as the final outcome. Left (Right): the estimated contrast function in stage 2 (1).}
\end{figure}

Finally, we estimate the optimal dynamic treatment regime in STAR*D with a different final outcome, 17-item Hamilton Rating Scale for Depression ($\text{HRSD}_{17}$), which was determined by independent, telephone-based interviewers with no knowledge of treatment assignment at the end of each level and was suggested in \cite{rush2004sequenced} as the primary outcome for research purposes (see Figure S5 in the supplement for the relationship between $\text{QIDS-C}_{16}$ and $\text{HRSD}_{17}$). That is, the final outcome is defined as  $Y=-(1-I_{\text{sim}})(\text{Hd}_2+\text{Hd}_3)/2-I_{\text{sim}}\text{Hd}_2$, where $\text{Hd}_j$ denotes the $\text{HRSD}_{17}$ at the end of the level $j$. Keeping the  specifications required in each method the same as before, we obtain the estimation results shown in Table \ref{tab:stard_par_hrsd} and Figure \ref{fig:stard_tree_hrsd}. The estimates of $\psi_{21}$ in Q- and A-learning are found to be significantly positive and the threshold values of $\Delta S_{2}$ for switching are closer to zero compared with those indicated in Table \ref{tab:stard_par}. Therefore, more participants are recommended to switch in stage 2 (level 3) when guided by achieving the best $\text{HRSD}_{17}$ instead of $\text{QIDS-C}_{16}$. As can be seen from this case study, different conclusions are obtained when the definition of `optimal' varies. It is therefore important to communicate with doctors to determine the final outcome which dictates the definition of optimal dynamic treatment regimes. Also, when reporting the obtained results, one should clarify the definition of ‘optimal'.

\begin{table}
\centering
\caption{Estimation results of $\psi_2=(\psi_{20},\psi_{21})$ and $\psi_1=(\psi_{10},\psi_{11})$ in STAR*D with the contrast function in stage $j$ specified as $\psi_{j0}+\psi_{j1}\Delta S_{j}$ in Q- and A-learning for $j=1,2$. The results under AIPWE are normalized via $\|\psi_1\|_2=1$ and $\|\psi_2\|_2=1$.} 
\begin{tabular}{crrrr}
  \hline \vspace{-2mm} \\ 
& \multicolumn{1}{c}{$\psi_{20}$} & \multicolumn{1}{c}{$\psi_{21}$}  & \multicolumn{1}{c}{$\psi_{10}$} & \multicolumn{1}{c}{$\psi_{11}$ }\\ \hline \vspace{-1mm}\\ 
 Working treatment-free & \multicolumn{2}{c}{$1+S_{2}+S_{1}+S_{0}$;} & \multicolumn{2}{c}{\quad $1+S_1+S_{0}$}\\
terms & & & \multicolumn{2}{c}{ \textcolor{blue}{\quad $1+S_1+\Delta S_{1}$}}\vspace{1mm}\\
Q &  $-1.5591$ (0.5686)&  $0.1938$ (0.1515) &  $-0.9078$ (0.3794)& $-0.0765$ (0.0872)\\ \vspace{2mm}
  &                    &             &\textcolor{blue}{$-0.9725$ (0.3959)} &\textcolor{blue}{$-0.2973$ (0.1337)}\\
A &  $-1.8081$ (0.5831) & 0.1609 (0.2056)  &  $-0.9868$ (0.4712)&$-0.1325$ (0.1271) \\ \vspace{2mm}
  &                    &             &\textcolor{blue}{$-0.9651$ (0.5234)} &\textcolor{blue}{$-0.1950$ (0.1677)}\\ 
 Working treatment-free &\multicolumn{2}{c}{$1+S_2+\Delta S_{2}$; } & \multicolumn{2}{c} {\quad $1+S_1+S_{0}$} \\
 terms & & & \multicolumn{2}{c} {\textcolor{blue}{\quad $1+S_1+\Delta S_{1}$}} \vspace{1mm}\\  
Q &  $-1.3489$ (0.4976)&  $-0.0220$ (0.2990) & $-0.9678$ (0.4307) & $-0.0913$ (0.0917) \\ \vspace{2mm}
 &                    &             &\textcolor{blue}{$-1.0199$ (0.4818)} &\textcolor{blue}{$-0.2696$ (0.1525)}\\
A &  $-1.5855$ (0.5961) & 0.0400 (0.3538) & $-1.0044$ (0.4712)& $-0.1374$ (0.1270)\\ \vspace{4mm}
 &                    &             &\textcolor{blue}{$-0.9838$ (0.5234)} &\textcolor{blue}{$-0.2002$ (0.1676)}\\
AIPWE &$-0.9782$ (0.1395)& $0.2075$ (0.3799)& $-0.9258$ (0.2336) & $0.3779$ (0.4085)
\\ \hline
\end{tabular}
\label{tab:stard_par}
\end{table}

\begin{table}
\centering
\caption{Estimation results of $\psi_2=(\psi_{20},\psi_{21})$ and $\psi_{10}$ in STAR*D with the contrast function in stage 2 specified as $\psi_{20}+\psi_{21}I\{\Delta S_{2}>3.7\}$ and that in stage 1 specified as $\psi_{10}$.} 
\begin{tabular}{crrr}
  \hline \vspace{-4mm} \\ 
& \multicolumn{1}{c}{$\psi_{20}$} & \multicolumn{1}{c}{$\psi_{21}$}  & \multicolumn{1}{c}{$\psi_{10}$}\\ \hline \vspace{-2mm}\\ 
Q &  $-1.8346$ (0.5784)&  $ 3.7326$ (1.6643) &  $-0.6970$ (0.3436)\\ 
A & $-2.0231$ (0.5667)&  $6.1023$ (2.6088)  &  $-0.7509$ (0.3503)
\\ \hline
\end{tabular}
\label{tab:stard_par_3.7}
\end{table}

\begin{table}
\centering
\caption{Estimation results of $\psi_2=(\psi_{20},\psi_{21})$ and $\psi_1=(\psi_{10},\psi_{11})$ in STAR*D with $\text{HRSD}_{17}$ as the final outcome.} 
\begin{tabular}{crrrr}
  \hline \vspace{-4mm} \\ 
& \multicolumn{1}{c}{$\psi_{20}$} & \multicolumn{1}{c}{$\psi_{21}$}  & \multicolumn{1}{c}{$\psi_{10}$}& \multicolumn{1}{c}{$\psi_{11}$}\\ \hline \vspace{-1mm}\\ 
Q &  $-0.5193$ (0.8711)&  $ 0.6545$ (0.2967) &   $-1.4555$ (0.6146) & $-0.2096$ (0.1103)\\ 
A & $-0.6342$ (0.9112)&  $0.5871$ (0.2590)  &  $-1.7988$ (0.6517) & $-0.2831$ (0.1548)\\
AIPWE & $-0.9095$ (0.3891) & 0.4158 (0.4937) & $-0.9930$ (0.3165)  & $-0.1181$ (0.3108)
\\ \hline
\end{tabular}
\label{tab:stard_par_hrsd}
\end{table}

\section{Discussion}\label{sec:discussion}
In this tutorial, we provide readers with a systematic, detailed but accessible introduction to optimal dynamic treatment regimes. We began by defining dynamic treatment regimes (DTRs) and what is an optimal DTR. We presented the assumptions required for the optimal DTR to be identifiable from the observed data and described some of the widely used methods for estimating it. These methods are classified as either indirect or direct depending on whether we first model the conditional mean or treatment contrast functions and then infer the optimal DTR from these models (see Algorithm 1); or we estimate the optimal DTR by targeting the maximization (assuming a larger outcome value is more desirable) of the expected potential final outcome over a specified class of candidate treatment regimes without any need for modelling the conditional mean or treatment contrast functions. 

Table \ref{tab:model_summary} summarised the discussed methods in terms of both model specification and estimation. Figure  \ref{fig:diagram} highlighted the generally increased protection against model misspecification as one moves from Q-learning and A-learning to causal tree and the direct methods based on inverse probability weighted and augmented inverse probability weighted estimators. Increased protection comes from the improved accuracy or greater flexibility when modelling the central model component using the observed data. However, decreased bias due to increased flexibility should be balanced with or traded-off against increased variability leading to potentially less generalisable decision rules.  Our simulation study explored the operating characteristics and performance of these methods with two or three decision points, one- or two-dimensional decision markers and with two sample sizes. The findings (under interpretable rules) are dependent on a number of factors including the degrees with which model misspecification and over fitting occur with respect to the central component at the different stages (and overall), and the specification of the working models for the auxiliary component(s) (e.g. propensity score or treatment-free terms). Our recommendation would be to apply a number of different methods in practice, with particular attention placed on choosing methods which return interpretable decision rules (e.g. linear or boolean decision rules). If contrast-based A-learning is one of the methods chosen then we would advocate for inclusion of judiciously specified non-null treatment-free terms as this can lead to improved efficiency due to reducing further the unexplained variability. If causal tree is chosen then some consideration should be given to whether the sample size is adequate to mitigate against over-fitting.

Prior medical knowledge is essential in determining the optimal dynamic treatment regimes. This tutorial is restricted to the case that we have sufficient information on which biomarkers/intermediate outcomes (i.e., $L_j$) should be incorporated when making a treatment decision rule in each stage and that $L_j$ is not of high dimension. Otherwise, variable selection would be necessary to ascertain the subset of biomarkers on which the treatment effect depends.  The fact that the contrast function (equivalently, the treatment effect function) rather than the conditional mean function is of most interest brings additional challenge in performing variable selection. See \cite{vansteelandt2012model} and \cite{powers2018some}  for strategies and discussions on this topic.

Non-regularity is an inherent issue in estimating optimal DTRs due to the max-operator involved in (\ref{2K}). Specifically, non-regularity refers to the case that the limiting distribution of an estimator depends in a non-uniform manner on the true value of the parameter, which is exactly the case for $\psi_{j}$ with $j\neq K$ \citep{laber2014dynamic}. That is, there is a subset of the parameter space where the limiting distribution of the estimator is no longer a mean-zero normal distribution. For example,  $\sqrt{n}\left(I\{\bar{X}\geq0\}\bar{X}-I\{\mu\geq 0\}\mu\right)$ converges to the positive part of a mean-zero normal random variable when $\mu=0$. When the parameter takes values close to the subset, the limiting distribution, though mean-zero normally distributed, provides a poor approximation to the finite sample distribution of the estimator.   It is therefore challenging to construct a valid confidence interval (at least the usual approach $\hat{\theta} \pm 1.96\times\text{SE} $ does not work well). We refer readers to \cite{laber2014dynamic} and Chapter 10 in \cite{tsiatis2019dynamic} for a detailed discussion of this issue and some developed methods for valid inference, such as utilizing the m-out-of-n bootstrap \citep{chakraborty2013inference}  to  construct confidence intervals, and incorporating individual selection in Q-learning to identify individuals whose treatment effects are close to zero \citep{song2015penalized}. Notably, the Bayesian framework  provides a natural approach to obviating the non-regular inferential challenges by accounting for the uncertainty in the optimal decision rules, i.e., by integrating over the posterior distribution of $\psi_j$ when making inference of $\psi_{j-1}$ \citep{murray2018bayesian}. 

\section*{Acknowledgments}
This work was supported by the United Kingdom Medical Research Council programme grant MC\_UU\_00002/2 and theme grant MC\_UU\_00040/02 (Precision Medicine) funding.

\section*{Data availability statement}
Real data subject to third party restrictions: The STAR*D data are available from the National Institute of Mental Health (NIMH) Data Archive. Restrictions apply to the availability of these data, which were used under license for this study. Simulated data can be generated with the R code provided in the supplementary R markdown file.

\label{Bibliography}
\bibliographystyle{apalike} % Use the "unsrtnat" BibTeX style for formatting the Bibliography

\bibliography{Bibliography} % The references (bibliography) information are stored in the file named "Bibliography.bib"

\end{document}